\documentclass[conference]{IEEEtran}

\usepackage[T1]{fontenc}
\usepackage{graphicx} % Required for including images
\usepackage{amsmath}
\usepackage{amssymb}
\usepackage{url}      % Required for formatting URLs
\usepackage{cite}     % Improves citation handling
\usepackage{listings} % Required for code blocks
\usepackage{xcolor}   % Required for defining colors in listings
\usepackage{hyperref} % Required for hyperlinks, load last
\usepackage{longtable} % For tables spanning multiple pages
\usepackage{booktabs}  % For professional quality horizontal rules in tables
\usepackage{array}     % For more advanced table column formatting

\hypersetup{
    colorlinks=true,
    linkcolor=blue,
    filecolor=magenta,
    urlcolor=cyan,
    citecolor=green,
    pdftitle={Agent Name Service (ANS): A Universal Directory for Secure AI Agent Discovery and Interoperability},
    pdfauthor={Ken Huang, Vineeth Sai Narajala, Idan Habler, Akram Sheriff},
    pdfkeywords={Agent Name Service (ANS), Agentic AI, Service Discovery, Public Key Infrastructure (PKI), Interoperability, Secure DNS, Formal Methods, Multi-Agent Systems (MAS)}, % Updated "Security" to "Secure DNS"
    breaklinks=true, % <--- ADD THIS LINE
}

\definecolor{codegreen}{rgb}{0,0.6,0}
\definecolor{codegray}{rgb}{0.5,0.5,0.5}
\definecolor{codepurple}{rgb}{0.58,0,0.82}
\definecolor{codeblue}{rgb}{0,0,1} % Define blue for keywords
\definecolor{backcolour}{rgb}{0.95,0.95,0.92}

\usepackage{academicons} % For \aiOrcid
\usepackage{xcolor} % academicons might need xcolor

\newcommand{\orcidicon}[1]{%
    \href{https://orcid.org/#1}{%
        \includegraphics[width=10pt]{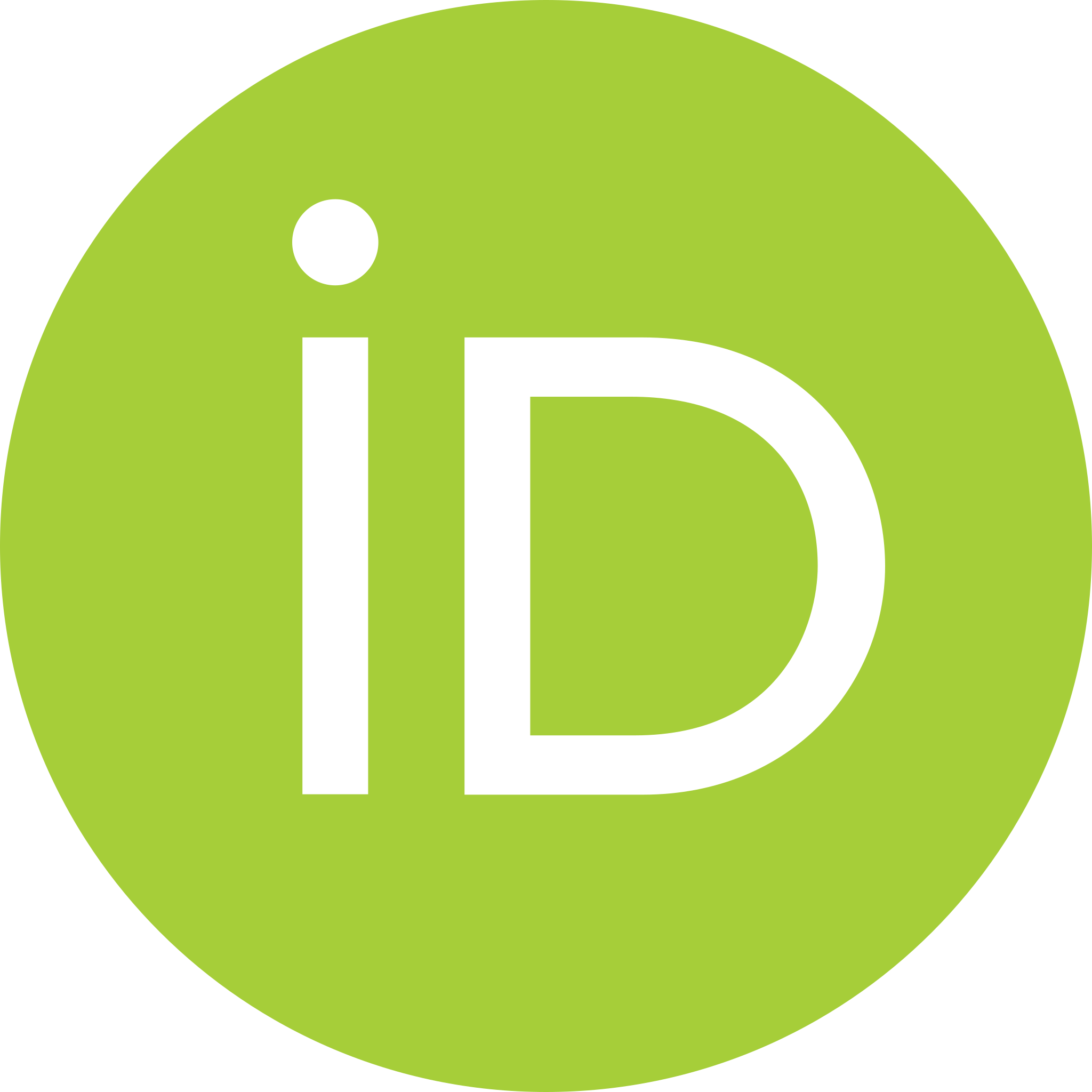}% % Assuming orcid_logo.png exists
    }%
}

\lstdefinestyle{jsonstyle}{
    backgroundcolor=\color{backcolour},
    commentstyle=\color{codegreen}, % JSON doesn't have standard comments, but keep for consistency
    keywordstyle=\color{codeblue},  % Style for keywords like true, false, null
    numberstyle=\tiny\color{codegray},
    stringstyle=\color{codepurple}, % Style for strings (keys and string values)
    basicstyle=\ttfamily\footnotesize,
    breakatwhitespace=false,
    breaklines=true,
    captionpos=b,
    keepspaces=true,
    numbers=left,
    numbersep=5pt,
    showspaces=false,
    showstringspaces=false,
    showtabs=false,
    tabsize=2,
    morestring=[b]", % Strings are delimited by "
    morekeywords={true, false, null}, % JSON keywords
}
\IEEEoverridecommandlockouts
\lstdefinestyle{pseudostyle}{
    backgroundcolor=\color{backcolour},
    commentstyle=\color{codegreen},
    keywordstyle=\color{blue}, % Keywords in blue
    numberstyle=\tiny\color{codegray},
    stringstyle=\color{codepurple},
    basicstyle=\ttfamily\footnotesize,
    breakatwhitespace=true, % Allow breaking at whitespace
    breaklines=true,
    captionpos=b,
    keepspaces=true,
    numbers=left,
    numbersep=5pt,
    showspaces=false,
    showstringspaces=false,
    showtabs=false,
    tabsize=2,
    morekeywords={Protocol, AgentID, agentCapability, Provider, Version, Extension, Cert, Sig, ANSName, Endpoint, String, Integer, Boolean, Set, VerifyCertChain, VerifySignature, Return, True, False, If, Else, Get, Check, Use, Hash, Compare, For, each, in, End, Query, Match, VersionNegotiation, GetAgentEndpointRecord, VerifyAgentEndpointRecord, IsVersionCompatible, Resolve, Parse, ERROR, Sort, by, OR, AND} % Add pseudocode keywords
}

\lstdefinestyle{verbatimstyle}{
    basicstyle=\ttfamily\footnotesize,
    breaklines=true,        % Enable line breaking
    breakatwhitespace=false, % Force breaks within words if necessary
    keepspaces=true,
    postbreak=\mbox{\textcolor{red}{$\hookrightarrow$}\space} % Add indicator for broken lines
}
\title{Agent Name Service (ANS): A Universal Directory for Secure AI Agent Discovery and Interoperability}

\author{
\IEEEauthorblockN{Ken Huang\textsuperscript{1} \thanks{\textsuperscript{1}This work is not related to the author’s position at DistributedApp.ai}}
\IEEEauthorblockA{\textit{Agentic AI Security} \\
\textit{DistributedApps.ai} \\
ken.huang@distributedapps.ai \orcidicon{0009-0004-6502-3673} }
\and
\IEEEauthorblockN{Vineeth Sai Narajala\textsuperscript{2} \thanks{\textsuperscript{2}This work is not related to the author’s position at Amazon Web Services.}}
\IEEEauthorblockA{\textit{Proactive Security} \\
\textit{Amazon Web Services} \\
vineesa@amazon.com \orcidicon{0009-0007-4553-9930}}
\and
\IEEEauthorblockN{Idan Habler\textsuperscript{3}  \thanks{\textsuperscript{3}This work is not related to the author’s position at Intuit}}
\IEEEauthorblockA{\textit{Adversarial AI Security reSearch (A2RS)} \\
\textit{Intuit} \\
idan\_habler@intuit.com \orcidicon{0000-0003-3423-5927}}
\and
\IEEEauthorblockN{Akram Sheriff\textsuperscript{4}  \thanks{\textsuperscript{4}This work is not related to the author’s position at Cisco Systems}}
\IEEEauthorblockA{\textit{AI Security} \\
\textit{Cisco Systems} \\
 isheriff@cisco.com \orcidicon{0000-0002-1606-7854} % Removed leading space
}}

\IEEEaftertitletext{\vspace{-0.5em}\begin{center}{\large \textbf{OWASP Gen AI Security Project - Agentic Security Initiative (ASI)}}\end{center}\vspace{0.5em}}

\begin{document}

\maketitle
\IEEEpeerreviewmaketitle % Add this for conference template peer review notice

\begin{abstract}
The proliferation of AI agents requires robust mechanisms for secure discovery. This paper introduces the Agent Name Service (ANS), a novel architecture based on DNS addressing the lack of a public agent discovery framework. ANS provides a protocol-agnostic registry mechanism that leverages Public Key Infrastructure (PKI) certificates for verifiable agent identity and trust. The architecture features several key innovations: a formalized agent registration and renewal mechanism for lifecycle management; DNS-inspired naming conventions with capability-aware resolution; a modular Protocol Adapter Layer supporting diverse communication standards (A2A, MCP, ACP, etc.); and precisely defined algorithms for secure resolution. We implement structured communication using JSON Schema and conduct a comprehensive threat analysis of our proposal. The result is a foundational agent directory service protocol addressing the core challenges of secure discovery and interaction in multi-agent systems, paving the way for future interoperable, trustworthy, and scalable agent ecosystems.
\end{abstract}


\section{Introduction}
\label{sec:introduction}

Agent-to-agent communication is expected to become a significant component of internet traffic, driving the need for reliable mechanisms enabling agents to discover, verify, and securely interact with one another. Traditional service discovery, notably the Domain Name System (DNS) \cite{RFC1035}, primarily maps human-readable names to network addresses and is insufficient for the dynamic, semantically rich, and security-sensitive environment of agentic AI. Enhancements like DNS-Based Service Discovery (DNS-SD) \cite{RFC6763} offer improvements but still fall short of the necessary agent capability granularity, identity verification, and lifecycle management required by autonomous agents. Furthermore, maintaining a trustworthy registry necessitates robust processes for agent registration and periodic renewal.

Several agent communication protocols are emerging to standardize interactions:
\begin{itemize}
    \item \textbf{Agent2Agent (A2A) Protocol \cite{Surapaneni2025, A2A2025}:} Developed by Google, providing a standardized protocol for inter-agent communication, aiming to bridge different agent frameworks.
    \item \textbf{Model Context Protocol (MCP) \cite{Anthropic2024, MCPSpec2025, Schmid2025}:} Focused on simplifying the integration of AI models with external tools and data sources.
    \item \textbf{Agent Communication Protocol (ACP) \cite{IBM2025}:} Designed to standardize how agents communicate, enabling automation, collaboration, UI integration, and developer tooling, evolving from initial MCP concepts.
\end{itemize}

This paper outlines the Agent Name Service (ANS), a framework for a protocol-agnostic Agentic AI Registry. ANS complements these emerging protocols by integrating Public Key Infrastructure (PKI) for identity and trust, defining structured communication via JSON Schema, incorporating DNS-like naming for discovery, establishing mechanisms for agent registration and renewal, and providing a formal specification of the protocol to enhance precision and implementability. ANS aims to provide a universal, secure directory service foundation for interoperable agent ecosystems.

\section{Related Work}
\label{sec:related_work}

Traditional service discovery, such as DNS \cite{RFC1035}, provides essential name-to-address resolution but lacks the semantic understanding and security features needed for agentic AI. DNS-SD \cite{RFC6763} adds local service discovery capabilities but doesn't address verifiable identity or complex agentCapability matching on a global scale.

Research in multi-agent systems (MAS) has explored various agent communication languages (ACLs), such as those defined by the Foundation for Intelligent Physical Agents (FIPA) \cite{FIPA2002}. While influential, these often lack standardized, built-in security mechanisms and universally adopted transport protocols suitable for the modern internet.

The emerging protocols represent significant advancements:
\begin{itemize}
    \item A2A \cite{Surapaneni2025, A2A2025} focuses on bridging agent ecosystems.
    \item MCP \cite{Anthropic2024, MCPSpec2025, Schmid2025} emphasizes dynamic discovery and integration of tools/data for AI models.
    \item ACP \cite{IBM2025} targets broader agent-to-agent communication needs, including delegation and orchestration.
\end{itemize}

Our work builds upon these efforts not by replacing them, but by providing a complementary, protocol-agnostic infrastructure layer. ANS differentiates itself by integrating PKI-based identity verification \cite{RFC5280} directly into the discovery and lifecycle management process, offering a universal registry mechanism that enhances trust and facilitates secure interaction across different protocol standards via a common discovery plan. Furthermore, the formalized specification of the ANS protocol ensures clarity and ease of implementation.

\section{Agent Registry Architecture}
\label{sec:architecture}

The proposed Agent Registry architecture provides a secure, interoperable platform for agent discovery and interaction, supporting multiple communication protocols through a modular design. Key components include:
\begin{itemize}
    \item \textbf{Requesting Agent:} The entity initiating the agent registration process, which could be an individual, organization, or automated system seeking to register a new agent or update existing agent information in the registry.
    \item \textbf{Agent Registry:} A potentially distributed database for storing ACEM (Agent Credential and Entitlement Management) and DID (Decentralized Identifier) related information. This registry encompasses agent capabilities, security policies, PKI certificates, protocol-specific metadata (via \texttt{protocolExtensions}), and registration/renewal timestamps, supporting a comprehensive framework for agent identity, authentication, and authorization.
    \item \textbf{Certificate Authority (CA):} A trusted entity issuing and managing X.509 digital certificates \cite{RFC5280} for agents, forming the root of trust.
    \item \textbf{Registration Authority (RA):} Verifies agent registration/renewal requests, interacts with the CA to issue certificates based on Certificate Signing Requests (CSRs), manages the agent lifecycle (registration, renewal, revocation), and validates the legal entity of the Requesting Agent. It enforces registry policies.
    \item \textbf{Protocol Adapter Layer:} Translates between the registry's internal representation and protocol-specific formats (details in Section \ref{sec:protocol_adapter}).
    \item \textbf{Request/Response Schema:} A protocol-agnostic JSON-based schema \cite{RFC7159} for registry interactions (discovery, registration, etc.), incorporating PKI data and allowing protocol-specific extensions (details in Section \ref{sec:schema}).
    \item \textbf{Agent Name Service (ANS):} Enables agent discovery using human-readable, structured names, coupled with agentCapability-based resolution (details in Section \ref{sec:ans_naming}).
\end{itemize}

The Protocol Adapter Layer translates between the registry's internal representation and protocol-specific formats. For example, consider an agent registering with the MCP. An MCP tool description might be represented as a JSON blob. Critically, the agent would need to be registered with ANS first and foremost. Therefore, imagine this tool is associated with the following ANSName: "\url{mcp://sentimentAnalyzer.textAnalysis.ExampleCorp.v1.0}". This would mean the MCP tool is now discoverable via the ANS. The MCP specific extension data itself might look like this:

\begin{lstlisting}[style=jsonstyle, caption={Example MCP Extension Data}, label={lst:mcp_extension_example}]
{
  "description": "Analyzes sentiment of text input.",
  "input_schema": {
    "type": "string",
    "description": "Text to analyze."
  },
  "output_schema": {
    "type": "object",
    "properties": {
      "sentiment": {
        "type": "string",
        "enum": ["positive", "negative", "neutral"]
      },
      "score": {
        "type": "number",
        "description": "Sentiment score (-1 to 1)."
      }
    }
  },
  "mcpEndpoint": "https://sentiment.example.com/analyze"
}
\end{lstlisting}
The MCP Adapter within the Protocol Adapter Layer would parse this JSON and map it to the registry's internal columns. This could involve:
\begin{itemize}
    \item Extracting information implicitly: since the ANSName is \url{mcp://sentimentAnalyzer.textAnalysis.ExampleCorp.v1.0}, this implicitly defines the:
    \begin{itemize}
        \item Protocol: \texttt{mcp}
        \item AgentID: \texttt{sentimentAnalyzer}
        \item agentCapability: \texttt{textAnalysis}
        \item Provider: \texttt{ExampleCorp}
        \item Version: \texttt{v1.0}
    \end{itemize}
    \item Storing the description ("Analyzes sentiment of text input.") in a dedicated description field within \texttt{protocolExtensions}.
    \item Serializing the \texttt{input\_schema} and \texttt{output\_schema} and storing them in a \texttt{protocolExtensions} column specific to MCP, allowing other MCP-aware agents to understand the tool's interface.
    \item The actual MCP endpoint \texttt{"https://sentiment.example.com/analyze"} would be stored within the \texttt{protocolExtensions}, under the key \texttt{mcpEndpoint}.
\end{itemize}
This normalization process allows the Agent Registry to store and query MCP-specific information in a protocol-agnostic way, while adhering to the ANSName structure for consistent identification and resolution.

\begin{figure*}[!t]
\centering
\includegraphics[width=0.95\linewidth]{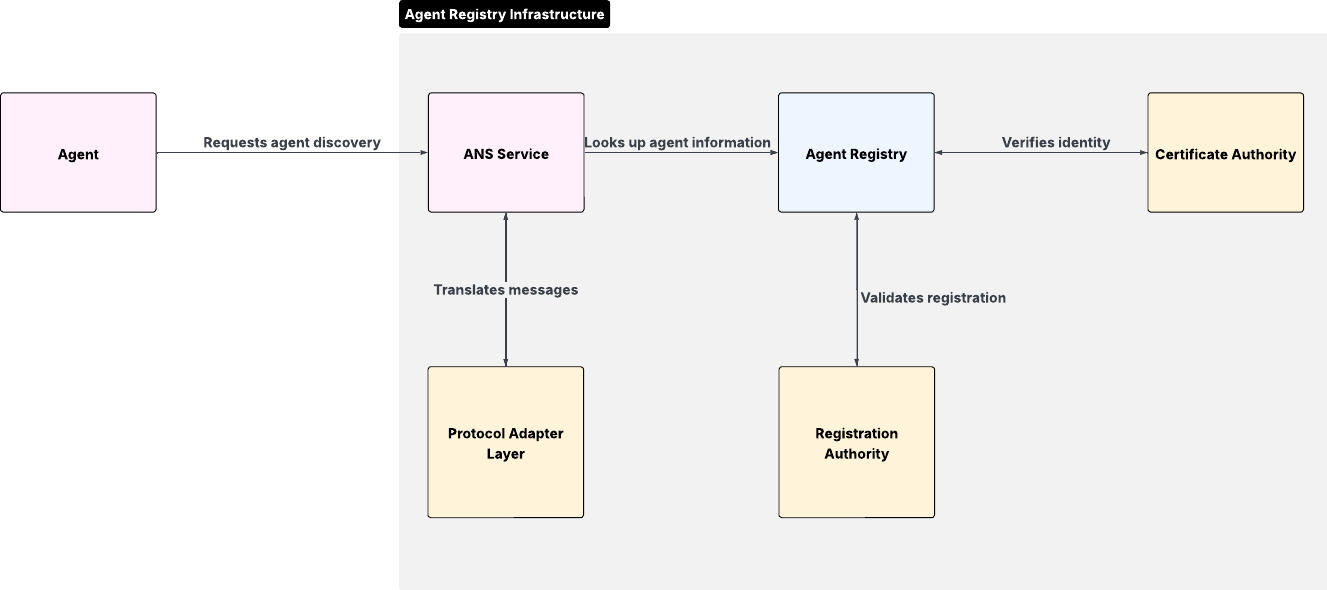} % Replace with actual image file
\caption{ANS Architecture. Illustrates the interaction between Agent, ANS Service, Agent Registry, CA, RA, and Protocol Adapter Layer.}
\label{fig:ans_architecture}
\end{figure*}

Figure \ref{fig:ans_architecture} illustrates the core components of the Agent Name Service (ANS) and how they interact.
\textit{Description of Figure 1: Imagine an Agent trying to find another agent. It starts by contacting the ANS Service, which acts like a central directory. The ANS Service relies on the Agent Registry, a specialized database containing information about all registered agents, including their capabilities, security policies, and contact details. The Agent Registry also interacts with a Certificate Authority (CA), which issues and manages digital certificates to verify agent identities, and a Registration Authority (RA), which validates new agents joining the system. To support different communication protocols, the ANS Service uses a Protocol Adapter Layer, which translates requests and responses into the appropriate format. All communication with ANS services is structured with JSON schemas. Together, these components form the Agent Registry Infrastructure that allows secure and reliable agent discovery.}

Table~\ref{tab:protocol_validation_pcol} outlines how Google A2A, Anthropic MCP, and IBM's ACP validate agents and enforce security within the ANS. Each protocol uses distinct adapter implementations and validation mechanisms—such as ZKPs (Zero-Knowledge Proofs), tool schema checks, and role-based controls—to ensure trusted identity, capability verification, and secure agent interactions. Zero-Knowledge Proofs (ZKPs) are cryptographic protocols that allow one party (the prover) to convince another party (the verifier) that a statement is true without revealing any information beyond the validity of the statement itself. These protocols maintain three essential properties: completeness (a valid proof always convinces an honest verifier), soundness (a false statement cannot be proven true except with negligible probability), and zero-knowledge (the verifier learns nothing beyond the statement's validity). ZKPs can be used for capability attestation by enabling agents to prove they possess certain abilities or skills without exposing the underlying data.

\begin{table*}[!t]
\centering
\small
\caption{Overview of how different protocols validate agents and enforce security within ANS.}
\label{tab:protocol_validation_pcol}
\newcommand{\colwidthA}{0.22\linewidth}
\newcommand{\colwidthB}{0.24\linewidth}
\newcommand{\colwidthC}{0.28\linewidth}
\newcommand{\colwidthD}{0.21\linewidth}
\begin{tabular}{|p{\colwidthA}|p{\colwidthB}|p{\colwidthC}|p{\colwidthD}|}
\hline
\multicolumn{1}{|>{\centering\arraybackslash}p{\colwidthA}|}{\textbf{PROTOCOL}} &
\multicolumn{1}{ >{\centering\arraybackslash}p{\colwidthB}|}{\textbf{ADAPTER IMPLEMENTATION}} &
\multicolumn{1}{ >{\centering\arraybackslash}p{\colwidthC}|}{\textbf{VALIDATION MECHANISM}} &
\multicolumn{1}{ >{\centering\arraybackslash}p{\colwidthD}|}{\textbf{SECURITY FEATURES}} \\
\hline
Google Agent (Agent 2 Agent) &
Native Implementation with Google SDK &
Agent ID card Integrity verification. &
Capability attestation with ZKP \\
\hline
Anthropic MCP &
Anthropic compliant adapter with extension validation &
Tool Identity / Tool Schema verification &
Resource access control \\
\hline
ACP (Agent Communication Protocol) &
IBM ACP Reference Implementation &
Role-based Agent Identity and capability enforcement. &
Delegation validation \\
\hline
\end{tabular}
\end{table*}

\subsection{Agent Registration and Renewal}
\label{sec:registration_renewal}

Maintaining registry integrity requires explicit lifecycle management:
\begin{itemize}
    \item \textbf{Registration:}
        \begin{enumerate}
            \item An agent submits a registration request (conforming to the defined JSON schema) including metadata, protocol details (within \texttt{protocolExtensions}), and a CSR.
            \item The RA validates the agent's identity and submitted information against registry policies (potentially involving automated checks or human review).
            \item The RA requests a certificate from the CA using the validated CSR.
            \item The issued certificate and agent information are stored in the Agent Registry.
        \end{enumerate}
    \item \textbf{Renewal:}
        \begin{enumerate}
            \item Agents periodically submit renewal requests before their registration or certificate expires.
            \item The RA verifies continued compliance with policies.
            \item The RA requests a new certificate from the CA.
            \item The agent's registration/renewal timestamp and potentially updated certificate are stored in the Registry.
        \end{enumerate}
    \item \textbf{Deregistration/Revocation:} Agents can be deregistered, or their certificates revoked (e.g., due to key compromise), removing or flagging their entry in the registry and invalidating their certificate via standard PKI mechanisms (CRL/OCSP \cite{RFC6960}).
\end{itemize}

\begin{figure*}[!t]
\centering
\includegraphics[width=0.95\linewidth]{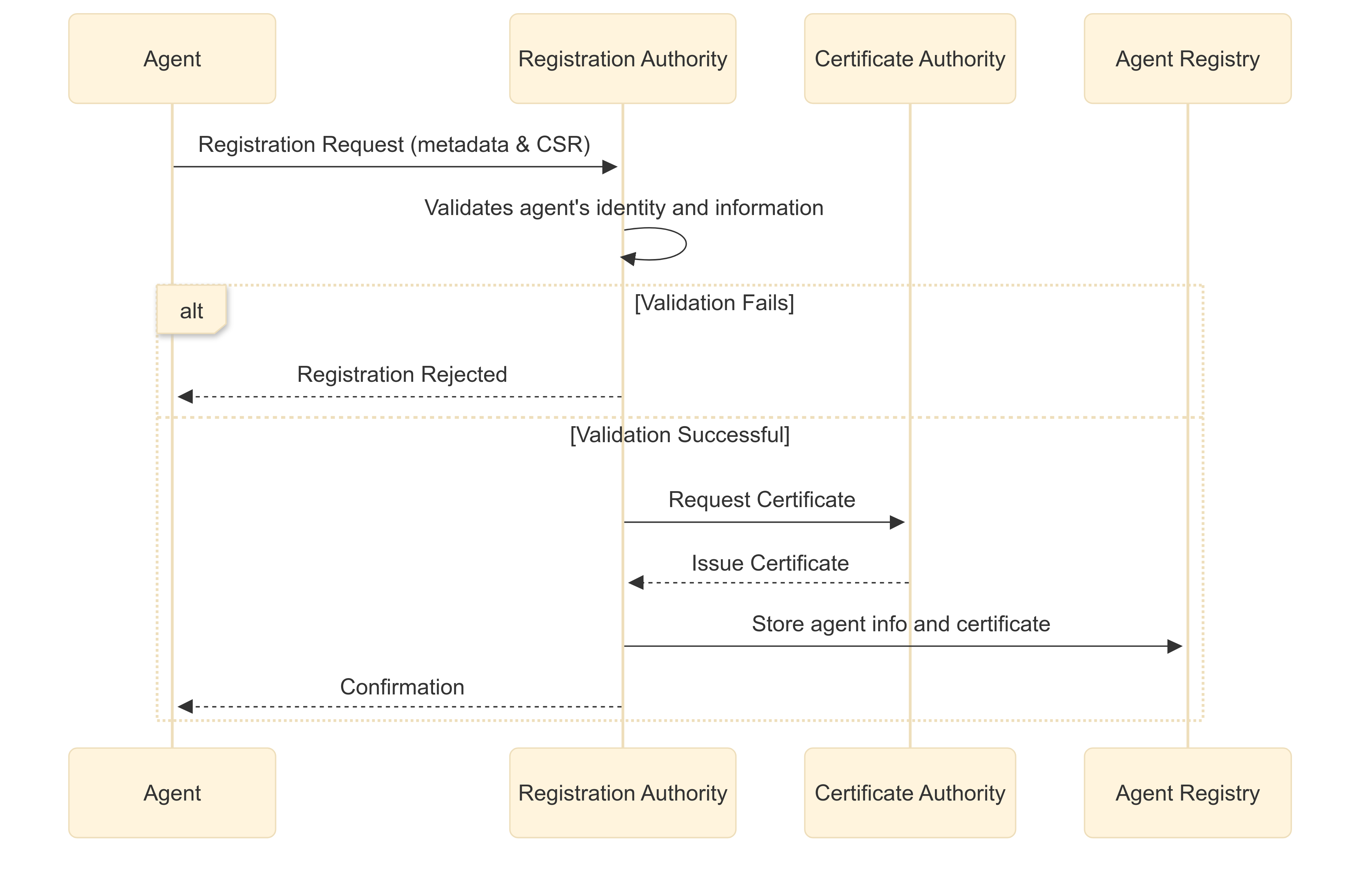} % Replace with actual image file
\caption{Agent Registration Process.}
\label{fig:registration_process}
\end{figure*}

Figure \ref{fig:registration_process} outlines the steps an agent takes to register with the Agent Name Service (ANS).
\textit{Description of Figure 2: The process begins when an Agent sends a Registration Request to the Registration Authority (RA). This request includes the agent's metadata (name, capabilities, etc.) and a Certificate Signing Request (CSR). The RA then Validates the agent's identity and information. If the validation Fails, the Registration is Rejected. If the validation is Successful, the RA requests a Certificate from the Certificate Authority (CA). The CA issues the certificate and sends it back to the RA. Finally, the RA stores the agent's information and certificate in the Agent Registry and sends a Confirmation back to the Agent, completing the registration process.}

\subsection{PKI Integration}
\label{sec:pki_integration}

PKI \cite{RFC5280} provides the foundation for trust. Each registered agent possesses a unique PKI key pair and a corresponding digital certificate issued by the CA via the RA.
\begin{itemize}
    \item \textbf{Identity Verification:} The certificate binds the agent's public key to its verified identity (e.g., its ANSName, organizational affiliation). Other agents can verify signatures made with the private key using the public key in the certificate, ensuring authenticity and integrity.
    \item \textbf{Trust Chain:} Certificates are validated against the trusted CA, establishing a chain of trust.
    \item \textbf{Lifecycle Management:} Certificate validity is tied to the registration/renewal cycle. Revoked certificates are handled using Certificate Revocation Lists (CRLs) or the Online Certificate Status Protocol (OCSP) \cite{RFC6960}.
    \item \textbf{Simplification:} While PKI management can be complex, the RA/CA interaction within the registry framework aims to streamline certificate issuance and renewal for agent developers compared to manual processes.
\end{itemize}

\subsection{ANS Protocol Notation}
\label{sec:protocol_notation}

We introduce the following notation for defining ANS elements and operations:

\subsubsection{Top Level Elements}
\textbf{\texttt{Protocol}:} Communication Protocol \par\medskip
\textbf{\texttt{AgentID}:} Agent Identifier \par\medskip
\textbf{\texttt{agentCapability}:} Agent Capability \par\medskip
\textbf{\texttt{Provider}:} Provider Name \par\medskip
\textbf{\texttt{Version}:} Version Number \par\medskip
\textbf{\texttt{Extension}:} Extension Metadata \par\medskip
\textbf{\texttt{Cert}:} Agent Certificate (X.509) \par\medskip
\textbf{\texttt{Sig}:} Digital Signature \par\medskip
\textbf{\texttt{ANSName}:} Agent Name Service Name \par\medskip
\textbf{\texttt{Endpoint}:} a resolvable endpoint \par\medskip

\subsubsection{Data Types}
\textbf{\texttt{String}:} Represents a sequence of characters. \par\medskip
\textbf{\texttt{Integer}:} Represents an integer number. \par\medskip
\textbf{\texttt{Boolean}:} Represents a boolean value (true or false). \par\medskip
\textbf{\texttt{Set<T>}:} Represents a set of elements of type T. \par\medskip

So, the top level elements have the following data types:
\textbf{\texttt{Protocol}:} \{\texttt{a2a}, \texttt{mcp}, \texttt{acp}, \dots\} % Enumerated set
\par\medskip
\textbf{\texttt{AgentID}:} \texttt{String} \par\medskip
\textbf{\texttt{agentCapability}:} \texttt{String} \par\medskip
\textbf{\texttt{Provider}:} \texttt{String} \par\medskip
\textbf{\texttt{Version}:} \texttt{String} (Semantic Versioning) \par\medskip
\textbf{\texttt{Extension}:} \texttt{String} \par\medskip
\textbf{\texttt{Cert}:} X.509 Certificate \par\medskip
\textbf{\texttt{Sig}:} Digital Signature \par\medskip
\textbf{\texttt{ANSName}:} \texttt{String} % Formatted string as defined below
\par\medskip
\textbf{\texttt{Endpoint}:} \texttt{String} % Network address, service binding
\par\medskip

\subsubsection{Verification Rules}

\textbf{Certificate Chain Verification}
\begin{lstlisting}[style=pseudostyle, caption={Certificate Chain Verification Algorithm}, label={lst:cert_chain_verify}, numbers=none]
VerifyCertChain (Cert, TrustedCA) -> Boolean:
  1. Get the certificate authority (CA) that signed the Cert
  2. Check for Certificate Revocation status of Cert via CRL or OCSP
  3. If Cert is Revoked Return False
  4. If CA == TrustedCA, Return True
  5. Else, recursively check CA Cert against TrustedCA
  6. If no trusted CA is found in the chain, Return False
\end{lstlisting}

\textbf{Digital Signature Verification}
\begin{lstlisting}[style=pseudostyle, caption={Digital Signature Verification Algorithm}, label={lst:sig_verify}, numbers=none]
VerifySignature (Data, Signature, PublicKey) -> Boolean:
  1. Use PublicKey to decrypt the Signature
  2. Hash Data using agreed upon function (e.g., SHA-256)
  3. Compare the decrypted signature to the Data hash
  4. If signatures are valid return True, otherwise return False
\end{lstlisting}

\subsection{Protocol-Agnostic Communication Schema}
\label{sec:schema}

We define JSON Schema \cite{RFC7159} documents for registry interactions (discovery requests/responses, registration/renewal requests/responses). This ensures structured, validated communication.

An example \texttt{AgentRegistrationRequest} schema is shown in Listing \ref{lst:agent_reg_req}.
\begin{lstlisting}[style=jsonstyle, caption={Example AgentRegistrationRequest JSON Schema}, label={lst:agent_reg_req}]
{
  "$schema": "http://json-schema.org/draft-07/schema#",
  "title": "AgentRegistrationRequest",
  "description": "Schema for Agent Registration Request",
  "type": "object",
  "properties": {
    "protocol": {
      "type": "string",
      "enum": ["a2a", "mcp", "acp"],
      "description": "Communication Protocol"
    },
    "agentID": {
      "type": "string",
      "description": "Unique Agent Identifier"
    },
    "agentCapability": {
      "type": "string",
      "description": "Primary Agent Capability"
    },
    "provider": {
      "type": "string",
      "description": "Name of the Provider"
    },
    "version": {
      "type": "string",
      "pattern": "^(0|[1-9]\\d*)\\.(0|[1-9]\\d*)\\.(0|[1-9]\\d*)(?:-((?:0|[1-9]\\d*|\\d*[a-zA-Z-][0-9a-zA-Z-]*)(?:\\.(?:0|[1-9]\\d*|\\d*[a-zA-Z-][0-9a-zA-Z-]*))*))?(?:\\+([0-9a-zA-Z-]+(?:\\.[0-9a-zA-Z-]+)*))?$",
      "description": "Semantic Versioning format"
    },
    "extension": {
      "type": "string",
      "description": "Extension Metadata"
    },
    "certificate": {
      "type": "object",
      "properties": {
        "subject": {
          "type": "string",
          "description": "Certificate Subject"
        },
        "issuer": {
          "type": "string",
          "description": "Certificate Issuer"
        },
        "pem": {
          "type": "string",
          "description": "PEM-encoded Certificate (strongly recommended to use a secure vault reference instead)",
          "readOnly": false
        }
      },
      "required": ["subject", "issuer", "pem"]
    },
    "protocolExtensions": {
      "type": "object",
      "description": "Protocol-specific data"
    }
  },
  "required": ["protocol", "agentID", "agentCapability", "provider", "version", "certificate"]
}
\end{lstlisting}

\begin{itemize}
    \item \textbf{Core Fields:} Include common elements like agent communication protocol types (\texttt{a2a}, \texttt{mcp}, \texttt{acp}, etc.), requesting/responding agent identifiers, timestamps, and PKI certificate details (subject, issuer, PEM representation - though referencing a secure vault is recommended for the PEM in production).
    \item \textbf{\texttt{protocolExtensions}:} A key field within the schema acts as a container for protocol-specific data (e.g., an A2A Agent Card \cite{securing_a2a}, MCP tool descriptions \cite{narajala2025enterprise}, ACP agent profiles). This allows the registry to store and query protocol-specific agentCapabilities while maintaining a common core schema.
    \item \textbf{Validation:} All interactions with the registry must be validated against these schemas. (See Section \ref{sec:request_response_schema} for more details on the schema structure).
\end{itemize}

\subsection{ANS Naming Structure and Resolution}
\label{sec:ans_naming}

ANS defines a robust, protocol-agnostic mechanism for naming and resolving agents across heterogeneous agentic environments. Its principal function is to establish a uniform \texttt{Endpoint} format that encodes identity, agentCapability, and contextual metadata for any given agent, irrespective of the underlying transport or runtime architecture. ANS ensures that both human-readable and machine-resolvable identifiers are preserved in a format designed to facilitate dynamic discovery, rigorous trust verification, secure communication, seamless service composition, and the representation of relationships between agents. A key motivation for ANS is to move beyond simple naming resolution to enable precise agentCapability discovery, which is not achievable with traditional systems like DNS. The design of ANS acknowledges that the agent's agentCapabilities are paramount for intelligent interactions, distinguishing it from simpler naming systems like DNS.

\subsubsection{Formal Naming Structure}
The \texttt{ANSName} is formally defined as a string constructed from the following components:
\begin{lstlisting}[style=verbatimstyle]
ANSName = Protocol "://" AgentID "." agentCapability
          "." Provider ".v" Version "." Extension
\end{lstlisting}
Where:
\begin{itemize}
    \item \texttt{Protocol} $\in$ \{\texttt{a2a, mcp, acp, ...}\}
    \item \texttt{AgentID}, \texttt{agentCapability}, \texttt{Provider}, \texttt{Version}, \texttt{Extension} are strings.
\end{itemize}

Constraints:
\begin{itemize}
    \item \texttt{Version} MUST adhere to Semantic Versioning standards.
    \item \texttt{AgentID}, \texttt{agentCapability}, \texttt{Provider} SHOULD be registered with a governance authority (similar to ICANN).
    \item \texttt{Extension} SHOULD be used for deployment-specific or provider-defined metadata, not for core identity. In the actual implementation, a registry of reserved tokens can be used to enhance security.
\end{itemize}
Example:
\begin{lstlisting}[style=verbatimstyle]
ANSName = "a2a://textProcessor.DocumentTranslation.AcmeCorp.v2.1.hipaa"
\end{lstlisting}

\subsubsection{Resolution}
The resolution mechanism in ANS is engineered to map a fully qualified \texttt{ANSName} to an actionable reference, such as a network address, service binding, or detailed metadata document (\texttt{Endpoint}). Resolution can be achieved through distributed lookups, local resolver caches, or enterprise-specific ANS gateways, providing deployment flexibility. Critically, ANS moves beyond simple name resolution to facilitate precise agentCapability discovery.

When an agent requires resolution, it queries the ANS service, a fundamental component of the Agent Registry infrastructure. The query includes the \texttt{ANSName} of the target agent and can incorporate optional agentCapability filters to refine the search.

\begin{figure*}[!t]
\centering
\includegraphics[width=0.8\linewidth]{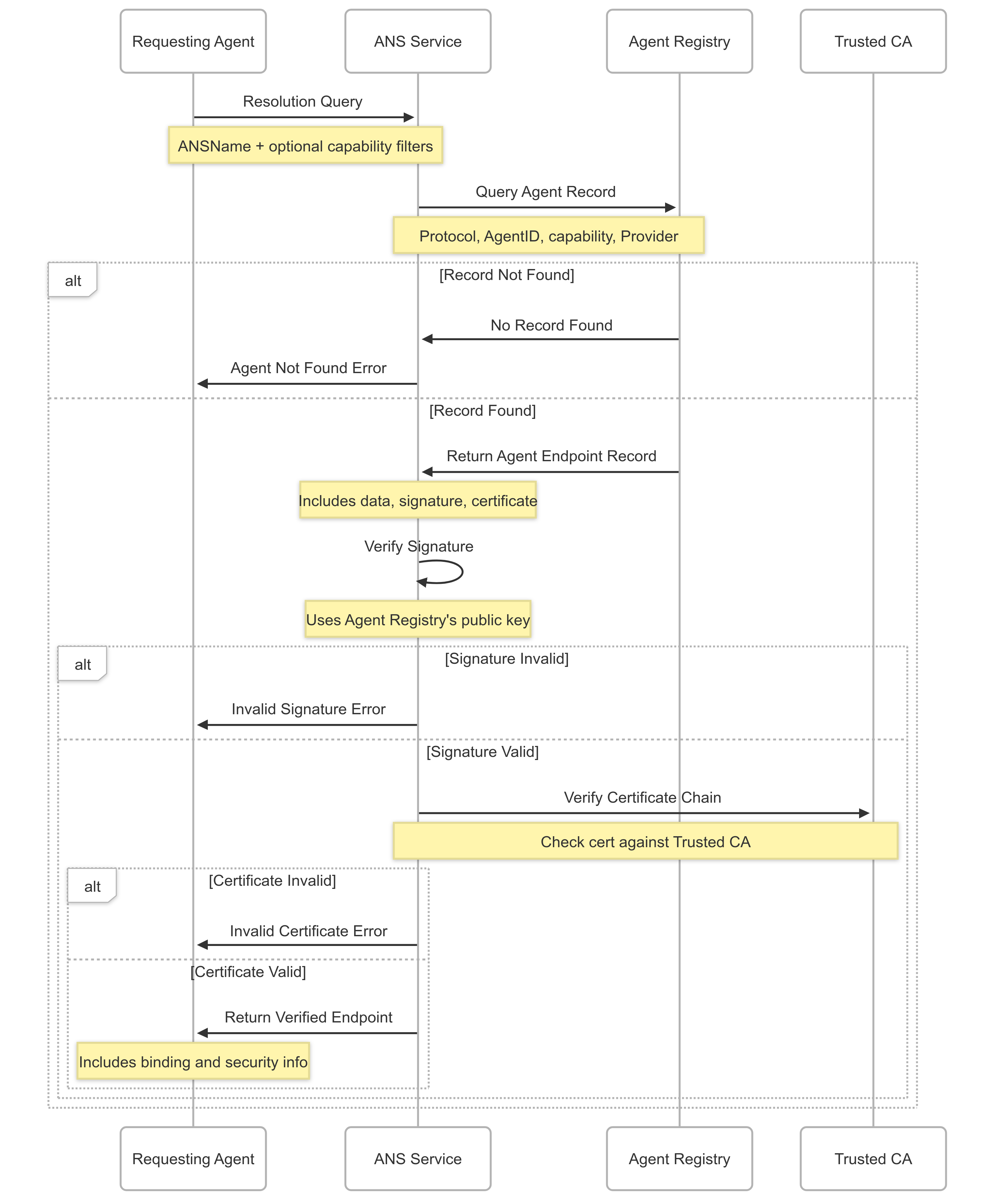} % Replace with actual image file
\caption{Agent Resolution Process.}
\label{fig:resolution_process}
\end{figure*}

Figure \ref{fig:resolution_process} depicts the agent resolution process.
\textit{Description of Figure 3: The process begins when an Agent sends a Resolution Query, containing the ANSName of the agent it wants to find, to the ANS Service. The ANS then Queries the Agent Registry for the corresponding agent record. If the Record is Not Found, the ANS returns an Agent Not Found Error. If the Record is Found, the ANS Verifies the returned Endpoint (which contains address and security information), including verifying target Agent’s signature and certificate. If the Verification Fails, an Invalid Endpoint Error is returned. If the Verification is Successful, the ANS returns the Endpoint to the Agent, enabling it to connect to the desired agent securely.}

\subsubsection{Formal Resolution Algorithm}
The ANS resolution algorithm takes an \texttt{ANSName} as input and returns a resolvable \texttt{Endpoint} or an error.

\begin{lstlisting}[style=pseudostyle, caption={ANS Resolution Algorithm}, label={lst:resolve_algo}]
Resolve(ANSName, RequestedVersionRange):
  1. Parse ANSName into Protocol, AgentID, agentCapability, Provider, Version, Extension
  2. Query Agent Registry for Agents with matching Protocol, AgentID, agentCapability, Provider
  3. If no match found:
     Return ERROR("Agent not found")
  4. If multiple matches found:
     Match = VersionNegotiation(Matches, RequestedVersionRange)
     if Match == ERROR("Incompatible Version")
       Return ERROR("Incompatible Version")
  5. EndpointRecord = GetAgentEndpointRecord(Match.AgentID)
  6. Valid = VerifyAgentEndpointRecord(EndpointRecord, TrustedCA)
  7. If valid == False
     Return ERROR ("Invalid Endpoint")
  8. Return Endpoint

GetAgentEndpointRecord () -> EndpointRecord
  // EndpointRecord: {data, signature, Cert}
  // Agent Registry implements GetAgentEndpointRecord to get records from the database.
  // GetAgentEndpointRecord enforces authentication and authorization with Agent Registry ACL.

VerifyAgentEndpointRecord (EndpointRecord, TrustedCA) -> Boolean:
  1. signatureValid = VerifySignature(EndpointRecord.data, EndpointRecord.signature, AgentRegistry.PublicKey)
  2. VerifySignature (Data, Signature, PublicKey) -> Boolean:
     1. Use PublicKey to decrypt the Signature
     2. Hash Data using agreed upon function
     3. Compare the decrypted signature to the Data hash
     4. If signatures are valid return True, otherwise return False
     // Verifying Signature is implemented by each language standard library, e.g., java.security.Signature
  3. If signature is invalid return ERROR ("Invalid Signature")
  4. certChainValid = VerifyCertChain(EndpointRecord.cert, TrustedCA)
  5. VerifyCertChain (Cert, TrustedCA) -> Boolean:
     1. Get the certificate authority (CA) that signed the Cert
     2. Check for Certificate Revocation status of Cert.
     3. If Revoked Return False
     4. If CA == TrustedCA, Return True
     5. Else, recursively check CA Cert against TrustedCA
     6. If no trusted CA is found in the chain, Return False
     // Certificate Chain validation, and Certificate Revocation Check are implemented via Library in standard language, e.g. Java: java.security.cert.CertPathValidator

VersionNegotiation(Matches, RequestedVersionRange):
  1. Sort Matches by Version (highest to lowest Semantic Version)
  2. For each Match in Matches:
  3.   If RequestedVersionRange == "*" OR IsVersionCompatible(Match.Version, RequestedVersionRange):
  4.     Return Match
  5.   End If
  6. End For
  7. Return ERROR("Incompatible Version")

IsVersionCompatible(AgentVersion, RequestedVersionRange) -> Boolean:
  // (Implement Semantic Version Range Compatibility Check here
  // Using existing library, e.g., https://github.com/npm/node-semver
  // 1. Attempt to parse requestedVersionRange as a SemVer range.
  // 2. If parsing fails, treat requestedVersionRange as a specific SemVer version.
  // 3. Check if agentVersion is satisfied by the requestedVersionRange.
  Return SemVer.satisfies(AgentVersion, RequestedVersionRange)
\end{lstlisting}

\textbf{IMPLEMENTATION NOTES:}
\begin{itemize}
    \item \textbf{Cacheability:} To ensure resolvers know when to re-validate EndpointRecords, the Agent Registry MUST include a Time-To-Live (TTL) value with each resolved Endpoint. The TTL indicates the number of seconds for which the EndpointRecord can be cached. A recommended default TTL is 300 seconds (5 minutes), but this value MAY be adjusted based on factors such as the volatility of the agent's configuration or the security policy of the Agent Registry. Resolvers MUST re-validate the EndpointRecord (by calling GetAgentEndpointRecord) after the TTL has expired.
    \item \textbf{Version Negotiation and Pre-release Tags:} When using \texttt{SemVer.satisfies} for version negotiation, pre-release tags (e.g., \texttt{-rc1}, \texttt{-beta}) MUST be considered to have lower precedence than the corresponding stable version. For example, version \texttt{1.0.0-rc1} would be considered lower precedence than \texttt{1.0.0}. This ensures that resolvers prefer stable versions over pre-release versions unless explicitly requested (e.g., by specifying a pre-release version range).
\end{itemize}

\subsubsection{Secure Resolution Implementation}
\begin{itemize}
    \item \textbf{Trust Anchor:} The trust anchor for ANS is the Agent Registry’s Certificate Authority (CA). The Agent Registry’s certificate is a public key certificate that is used to verify the digital signatures of the Agent Registry’s responses. The Agent Registry’s certificate must be trusted by all agents that use ANS.
    \item \textbf{Digital Signatures:} Digital signatures are used to ensure the integrity and authenticity of Agent Registry’s responses. The Agent Registry’s responses are digitally signed using the Agent Registry’s private key. Clients, upon receiving a response, verify the signature using the corresponding public key, confirming the integrity and authenticity of the data.
    \item \textbf{DNSSEC-like Security:} Consider the implementation of the Domain Name System Security Extensions (DNSSEC)-like mechanisms to validate the chain of trust. DNSSEC can increase the risk and amplify the effects of denial of service attacks on the infrastructure. DNSSEC also increases the number of DNS query responses because of the crypto fields that are used to verify records properly. This means that high-volume responses enable attackers with greater attack volume against a zone than they could if DNSSEC were not in place. Therefore, a careful evaluation of the threat model and the potential for amplification attacks is crucial before implementing DNSSEC-like security measures. Mitigation strategies such as rate limiting, traffic filtering, and anycast deployment should be considered to protect the Agent Registry infrastructure from potential DoS attacks.
    \item \textbf{Certificate Revocation:} Implement a robust mechanism for certificate revocation. If the Agent Registry's private key is compromised, the corresponding certificate must be revoked immediately to prevent attackers from using the compromised key to sign malicious responses. Use standard certificate revocation methods such as Certificate Revocation Lists (CRLs) \cite{RFC5280} or the Online Certificate Status Protocol (OCSP) \cite{RFC6960}.
    \item \textbf{Threat Modeling:} Perform ongoing threat modeling to identify potential vulnerabilities in the secure resolution mechanism. This will help to proactively address security concerns and ensure the ongoing security of ANS.
\end{itemize}

\begin{figure*}[!t]
\centering
\includegraphics[width=0.95\linewidth]{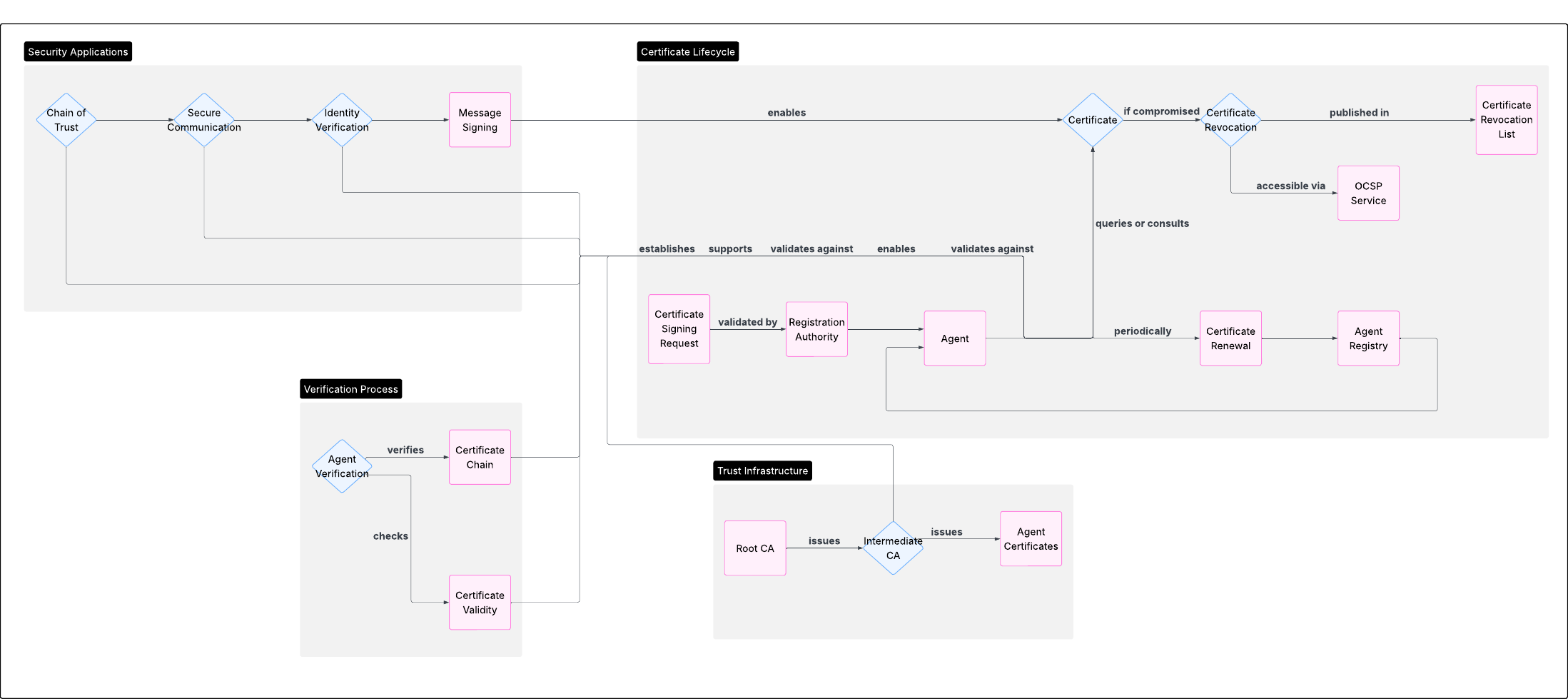} % Replace with actual image file
\caption{Full end-to-end flow (Registration, Resolution, Interaction).}
\label{fig:full_flow}
\end{figure*}

\begin{figure*}[!t]
\centering
\includegraphics[width=0.95\linewidth]{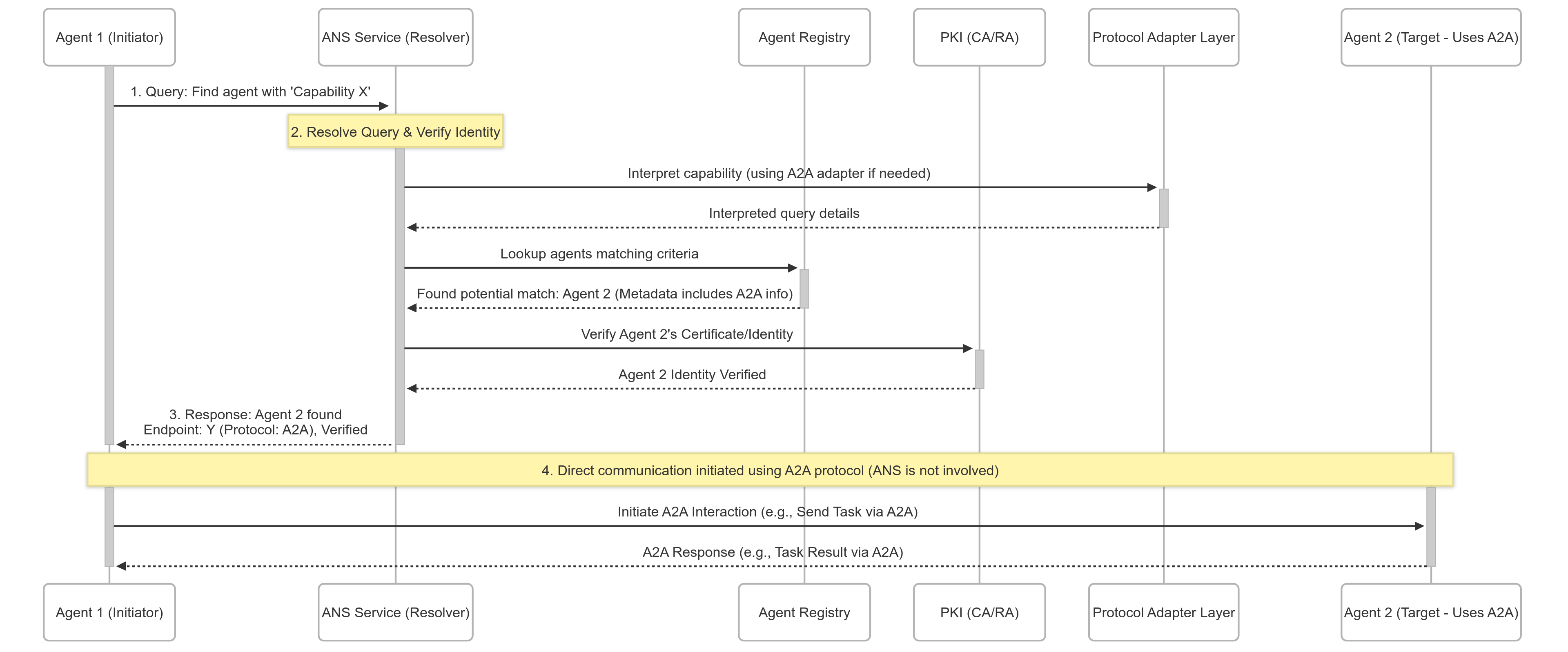}
\caption{ANS Resolution Sequence as a Pre-Communication Step.}
\label{fig:ans_resolution_sequence} % Make sure this label is unique
\end{figure*}

The overall end-to-end flow within the ANS ecosystem, covering registration, secure resolution, and following interactions, is illustrated Figure \ref{fig:full_flow}. To demonstrate the transition from discovery to direct communication in more detail, Figure \ref{fig:ans_resolution_sequence} presents the specific sequence where ANS facilitates secure connection setup as a distinct pre-communication phase. Steps 1-3 in the Figure illustrates the initiating agent leverages the ANS service, registry, and PKI to securely resolve and verify the target agent's endpoint. Following successful resolution via ANS, the agents proceed with direct communication using their native protocol (Step 4), where ANS is no longer directly involved.

\subsection{ANS Challenges and Governance}
\label{sec:challenges_governance}

Deploying ANS involves addressing key challenges:
\begin{itemize}
    \item \textbf{Naming Collisions/Squatting:} Ensuring uniqueness requires a managed registration process for names, particularly \texttt{<AgentID>}, \texttt{<agentCapability>}, and \texttt{<Provider>} segments. A governance model, potentially similar to ICANN for DNS, might be needed to manage top-level agentCapabilities and provider identifiers.
    \item \textbf{Scalability:} Supporting potentially billions of agents requires scalable registry storage (e.g., distributed databases, NoSQL) and efficient resolution mechanisms (e.g., distributed hash tables (DHTs), caching layers, geographically distributed resolution points).
    \item \textbf{Governance:} Establishing policies for name allocation, dispute resolution, operational practices, and managing the trust infrastructure (CAs, RAs) is crucial for long-term stability and trustworthiness.
\end{itemize}

\subsection{Agent Identity}
\label{sec:agent_identity}

Agent identity within the ANS framework can include the following:
\begin{itemize}
    \item \textbf{Cryptographic Identity:} The agent's PKI certificate provides a verifiable, CA-signed cryptographic identity.
    \item \textbf{Logical Identity:} The \texttt{ANSName} provides a human-readable, structured identifier conveying agentCapability, provider, etc.
    \item \textbf{Protocol-Specific Identity:} Agents may have identities within their native protocols (A2A Agent Card ID, MCP tool identifiers, ACP agent URIs), stored within the \texttt{protocolExtensions}.
    \item \textbf{Verifiable Claims:} The registry could support attaching digitally signed attestations (e.g., compliance certifications, capability endorsements) to agent profiles.
    \item \textbf{Identity linkage:} ANS ecosystem leverages the structured naming convention to establish relationships between agents. The core principle is that an agent's \texttt{ANSName} serves as a unique and resolvable identifier, allowing other agents or the system itself to reference it. The Agent Registry, upon recognizing this request, not only returns the target agent's details (binding, metadata, certificates) but also automatically resolves the linked agents, effectively materializing the relationship and providing all necessary information for secure and informed interaction. This automated relationship discovery, built upon the foundation of uniquely identifiable and resolvable agent names, significantly simplifies the orchestration and coordination of complex multi-agent systems.
\end{itemize}

\textbf{Agent Card Validation:} The integrity of Agent Cards within the ANS ecosystem is verified through cryptographic methods, utilizing the relational patterns between agents. This process ensures that capability declarations are validated against organizational policies. Additionally, endpoint URL structures are enforced to comply with security standards such as TLS and proper domain constraints. The Agent Registry oversees and connects these identity components, facilitating verification through challenge-response protocols that rely on the agent's private key. Both the Requesting Agent and the Registration Authority (RA) play crucial and distinct roles in this validation process.

\textbf{Requesting Agent Responsibility:} The Requesting Agent has a primary and ongoing responsibility for validating the Agent Card before every interaction. This includes:
\begin{itemize}
    \item Cryptographic Verification: Verifying the Agent Card's digital signature to ensure it hasn't been tampered with.
    \item Capability Alignment: Confirming that the agent's stated capabilities are actually what the Requesting Agent expects and needs for the intended interaction. This might involve checking specific input/output schemas or testing the agent's performance on sample tasks before relying on it for critical operations.
\end{itemize}
The Requesting Agent's validation is not a one-time event; it's a continuous process that ensures the agent remains trustworthy for each specific interaction. A failure to properly validate an Agent Card could expose the Requesting Agent to significant security risks.

\textbf{Registration Authority (RA) Responsibility:} The RA performs a foundational validation of the Agent Card during the agent registration and renewal processes. This includes:
\begin{itemize}
    \item Signature Verification: Verifying the Agent Card's signature and the validity of the associated certificate.
    \item Policy Adherence: Ensuring the agent's claimed capabilities and operational practices comply with broader registry policies and legal requirements.
    \item Legitimacy Checks: Performing checks to confirm the identity and legitimacy of the agent's owner (e.g., domain validation, organizational checks).
\end{itemize}
The RA's validation provides a baseline level of trust, but it does not replace the need for the Requesting Agent to perform its own, more context-specific validation.

Additionally, endpoint URL structures are enforced to comply with security standards such as TLS and proper domain constraints. The Agent Registry oversees and connects these identity components, facilitating verification through challenge-response protocols that rely on the agent's private key.

\textbf{Agent Capability Attestation:} The AI agent's identity and claimed capabilities are authenticated through zero-knowledge proof methods. Specifically, ZKPs can be employed to allow an agent to prove that it possesses certain capabilities (e.g., access to specific data, the ability to perform a certain computation) without revealing how it possesses those capabilities or the underlying data itself. For example, an agent might use a ZKP to prove it has access to a database containing sensitive patient information without revealing the specific query it will use or any of the patient data. This involves the agent constructing a proof, based on its private knowledge and the claimed capabilities, that can be verified by the Agent Registry (or another agent) using only publicly available information. The verifier gains assurance that the agent possesses the claimed capabilities without learning any sensitive information about the agent's internal state or data. During runtime, capabilities are dynamically validated as part of the resolution process. To further enhance real-time verification, challenge-response mechanisms are employed.

\textit{Challenge-Response Example:} Imagine an agent claims to be able to perform "Sentiment Analysis" with a certain accuracy.
\begin{itemize}
    \item The Agent Registry (or a verifying agent) sends the claimed "Sentiment Analysis" agent a specific challenge: a piece of text with a known sentiment.
    \item The "Sentiment Analysis" agent processes the text and returns its sentiment classification (positive, negative, neutral) and a confidence score.
    \item The Agent Registry (or verifying agent) compares the agent's response to the known sentiment and the claimed accuracy.
    \item If the response is correct and the confidence score aligns with the agent's claimed accuracy, the agent's capability is considered validated (for that specific challenge).
    \item If the response is incorrect or the confidence score is significantly lower than the claimed accuracy, the agent's claimed capability is called into question and further challenges or even revocation of the agent's registration might be triggered.
\end{itemize}
This challenge-response process can be repeated periodically or triggered based on certain events (e.g., a change in the agent's code, a security alert). The challenges can be designed to test various aspects of the agent's claimed capabilities, ensuring that it continues to function as expected over time. The Agent Registry maintains a history of challenge-response results to track the agent's performance and reliability.

By combining ZKPs for initial capability attestation with challenge-response mechanisms for ongoing validation, the ANS provides a robust framework for ensuring the trustworthiness of AI agents.

\textbf{Authentication Enforcement:} The process involves validating the OAuth 2.0 flow to ensure the legitimacy of authorization tokens, verifying mTLS certificates to confirm alignment with the registered agent's identity, and checking JSON Web Tokens (JWTs) to ensure their signatures and claims are accurate and properly authenticated.

\textbf{Agent Identity Module Examples:}
The Agent Identity module implements resource access control through capability-based security. Below are examples for A2A and MCP protocols:
\begin{lstlisting}[style=jsonstyle, caption={A2A Capability Verification Example}, label={lst:a2a_cap_ver_example}]
{
  "a2aCapabilityVerification": {
    "capabilityVerification": {
      "proofMechanism": "ZKP",
      "verificationCircuit": {
        "constraints": [
          "agent.hasCapability(c) AND agent.isAuthorized(c)",
          "agent.certificate.isValid() AND agent.certificate.notRevoked()"
        ],
        "proofGeneration": "Groth16",
        "verificationKey": "0x4a8f..."
      }
    },
    "rateLimit": {
      "algorithm": "TokenBucket",
      "refillRate": "100/s",
      "burstCapacity": 500,
      "perCapability": true
    }
  }
}
\end{lstlisting}

\begin{lstlisting}[style=jsonstyle, caption={MCP Agent Identity Example}, label={lst:mcp_agent_id_example}]
{
  "mcpAgentIdentity": {
    "resourceAccessControl": {
      "model": "RBAC+ABAC",
      "policyDecisionPoint": {
        "engine": "OPA",
        "evaluationMode": "distributed"
      },
      "contextAttributes": [
        "agent.role",
        "resource.classification",
        "time.window",
        "operation.sensitivity"
      ]
    },
    "toolRegistration": {
      "sandboxValidation": {
        "environment": "gVisor",
        "runtime": "V8Isolate",
        "memoryLimit": "256MB",
        "cpuQuota": "0.5",
        "networkPolicy": "DENY_ALL"
      }
    }
  }
}
\end{lstlisting}

The Agent Registry manages and links these identity facets, enabling verification via challenge-response protocols using the agent's private key.

\section{Request/Response Schema for ANS Name Resolution}
\label{sec:request_response_schema}

The following core JSON Schema defines the structure for Agent Capability requests and responses.

\textbf{AgentCapabilityRequest Schema:}
\begin{lstlisting}[style=jsonstyle, caption={AgentCapabilityRequest JSON Schema}, label={lst:agent_cap_req}]
{
    "$schema": "http://json-schema.org/draft-07/schema#",
    "title": "AgentCapabilityRequest",
    "description": "Schema for Agent agentCapability Request",
    "type": "object",
    "properties": {
        "requestType": {
            "type": "string",
            "enum": [
                "resolve"
            ],
            "description": "Type of request"
        },
        "protocol": {
            "type": "string",
            "enum": [
                "a2a",
                "mcp",
                "acp"
            ],
            "description": "Communication Protocol"
        },
        "agentID": {
            "type": "string",
            "description": "Unique Agent Identifier"
        },
        "agentCapability": {
            "type": "string",
            "description": "Primary Agent Capability"
        },
        "provider": {
            "type": "string",
            "description": "Name of the Provider"
        },
        "version": {
            "type": "string",
            "pattern": "^(0|[1-9]\\d*)\\.(0|[1-9]\\d*)\\.(0|[1-9]\\d*)(?:-((?:0|[1-9]\\d*|\\d*[a-zA-Z-][0-9a-zA-Z-]*)(?:\\.(?:0|[1-9]\\d*|\\d*[a-zA-Z-][0-9a-zA-Z-]*))*))?(?:\\+([0-9a-zA-Z-]+(?:\\.[0-9a-zA-Z-]+)*))?$",
            "description": "Semantic Versioning format"
        },
        "extension": {
            "type": "string",
            "description": "Extension Metadata"
        }
    },
    "required": [
        "requestType",
        "protocol",
        "agentID",
        "agentCapability",
        "provider",
        "version"
    ]
}
\end{lstlisting}

\textbf{AgentCapabilityResponse Schema:}
\begin{lstlisting}[style=jsonstyle, caption={AgentCapabilityResponse JSON Schema}, label={lst:agent_cap_resp}]
{
    "$schema": "http://json-schema.org/draft-07/schema#",
    "title": "AgentCapabilityResponse",
    "description": "Schema for Agent agentCapability Response",
    "type": "object",
    "properties": {
        "Endpoint": {
            "type": "string",
            "description": "Agent address (e.g., a2a://translatorBot.DocumentTranslation.exampleCorp.v1.2.3.secure)"
        },
        "signature": {
            "type": "string",
            "description": "signature"
        },
        "cert": {
            "type": "string",
            "description": "PEM-encoded Certificate (strongly recommended to use a secure vault reference instead)",
            "readOnly": false
        }
    },
    "required": [
        "Endpoint",
        "signature",
        "cert"
    ]
}
\end{lstlisting}

Key points regarding schemas:
\begin{itemize}
    \item Use a JSON Schema validator library for enforcement.
    \item Pay attention to required fields.
    \item Validate all incoming/outgoing messages against this schema.
    \item Handle validation errors gracefully.
    \item Monitor evolving standards (A2A, MCP, ACP) and update schemas accordingly.
\end{itemize}

\section{Protocol Adapter Layer}
\label{sec:protocol_adapter}

The Protocol Adapter Layer enables the registry to support diverse agent communication protocols without being tightly coupled to any single one. It acts as an intermediary between the registry's core logic/storage and the specific requirements of each protocol.

\begin{itemize}
    \item \textbf{Modularity:} Adapters are implemented as distinct modules (e.g., plugins).
    \item \textbf{Functionality:} Each adapter understands how to:
        \begin{itemize}
            \item Parse protocol-specific metadata (e.g., from an A2A Agent Card) and map relevant parts into the registry's internal representation (especially within \texttt{protocolExtensions}).
            \item Extract information from the registry record to answer protocol-specific discovery queries.
            \item Potentially handle protocol-specific aspects of registration or validation if needed.
        \end{itemize}
    \item Each adapter is responsible to securely implement features defined by each protocols. Some protocols have built in security, such as signed messages.
    \item \textbf{Translation:} Adapters primarily focus on metadata translation for discovery and registration, not on real-time message translation between protocols during agent interaction. They help agents find each other and verify identity; subsequent communication typically uses the agents' shared native protocol.
    \item \textbf{Security aspect:}
        \begin{itemize}
            \item All adapter implementations must use secure, updated libraries that are commonly used to implement the supported protocols.
            \item All adapter implementations must follow the security guidance by each protocol.
            \item Each adapter parses untrusted blobs; mandate memory-safe languages (Rust/Go) and formal test suites.
        \end{itemize}
\end{itemize}

\subsection{A2A Protocol Adapter}
\begin{itemize}
    \item Parses/stores A2A Agent Card information within \texttt{protocolExtensions}.
    \item Enables discovery based on A2A agentCapabilities advertised in the card.
    \item Facilitates finding A2A Endpoints listed in the card.
\end{itemize}

\subsection{MCP Adapter}
\begin{itemize}
    \item Parses/stores MCP Tool and Resource descriptions within \texttt{protocolExtensions}.
    \item Enables discovery of agents offering specific MCP tools/resources.
    \item Facilitates finding MCP-compliant Endpoints.
\end{itemize}

\subsection{ACP Adapter}
\begin{itemize}
    \item Parses/stores ACP agent profiles and agentCapability advertisements within \texttt{protocolExtensions}.
    \item Supports discovery based on ACP roles or agentCapabilities.
    \item May assist in bootstrapping ACP delegation or orchestration workflows by providing initial agent references.
\end{itemize}

\begin{figure*}[!t]
\centering
\includegraphics[width=0.95\linewidth]{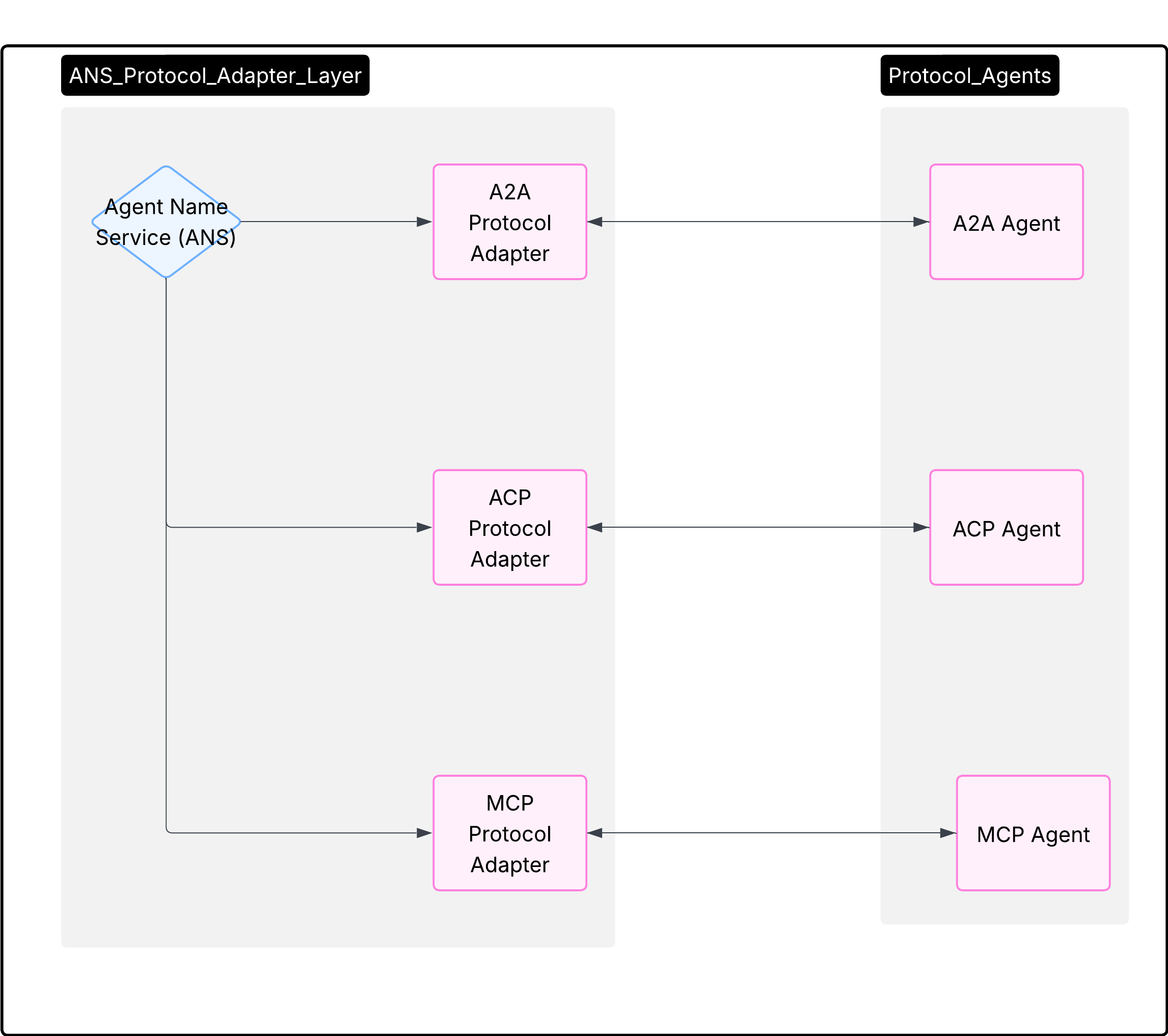} % Replace with actual image file
\caption{Protocol Adapters connecting the core registry to different agent protocols (A2A, MCP, ACP).}
\label{fig:protocol_adapters}
\end{figure*}

Figure \ref{fig:protocol_adapters} illustrates the concept of protocol adapters.

\subsection{Extension Points}
Adding support for a new protocol involves creating a new adapter module that implements the required mapping and discovery logic interfaces defined by the registry framework. This ensures the registry can evolve with the agent ecosystem.

\subsection{Cross-Protocol Interoperability Limits}
The registry primarily enables discovery and identity verification across protocols. An A2A agent can discover an agent advertising MCP tools, verify its identity via PKI, and potentially interact if it also understands MCP or uses a gateway. ANS does not automatically translate A2A tasks into MCP requests. It provides the foundational trust and location information, but deeper semantic interoperability often requires additional mechanisms or multi-protocol support within the agents themselves.

\subsection{Protocol Adapter API Definition}
Protocol Adapters are implemented as plugins that conform to the following interface (expressed in a language-agnostic way; adapt to your chosen implementation language):

\begin{lstlisting}[style=pseudostyle, caption={Protocol Adapter Interface}, label={lst:adapter_interface}, numbers=none, language=Java] // Using Java-like syntax for interface
interface ProtocolAdapter {
  // Identifies the protocol supported by this adapter (e.g., "a2a", "mcp", "acp")
  String getProtocol();

  // Parses protocol-specific metadata from the protocolExtensions and maps it to an internal registry format.
  // Returns a map of key-value pairs representing the extracted data.
  Map<String, Object> parseMetadata(Object protocolExtensions);

  // Creates a protocol-specific discovery response based on the registry record.
  Object createDiscoveryResponse(Map<String, Object> registryRecord);

  // Handles protocol-specific validation during the registration process
  Boolean validateRegistration(AgentRegistrationRequest request);
}
\end{lstlisting}

Important Considerations:
\begin{itemize}
    \item This is a basic interface. More complex scenarios might require additional methods (e.g., for handling updates, deletions, etc.).
    \item The \texttt{Map<String, Object>} type should be replaced with more specific data structures based on your chosen implementation language and registry data model.
    \item Error handling should be implemented using exceptions or appropriate return codes.
\end{itemize}

\subsection*{SUMMARY OF ANS FUNCTIONAL LAYERS}
\begin{table*}[!htbp]
\centering
\caption{Summary of ANS Functional Layers}
\label{tab:ans_functional_layers}
\small
\begin{tabular}{|p{0.25\linewidth}|p{0.25\linewidth}|p{0.4\linewidth}|}
\hline
\textbf{ANS Component} & \textbf{ANS Functional Layer} & \textbf{JSON Schema Implementation} \\
\hline
Policy Enforcement Engine & Request Processing & Forward-chaining rule processor with sequential evaluation of Agent's certs. \\
\hline
PKI Governance & Security Infrastructure & Hierarchical PKI with Agent specific Cert Validation \& Integration. \\
\hline
A2A Protocol Engine & Protocol Adapter & Zero-knowledge proof capability verification \\
\hline
MCP Protocol Engine & Protocol Adapter & RBAC+ABAC access control with OPA \\
\hline
Consensus Engine & Distributed Governance & Analyzing Agent Signatures \\
\hline
ANS Audit Trail System & Compliance Layer & Deterministically generate unique Agent IDs (UUIDv5 based on PKI public key hash) \\ % Note: Original PDF had "**Ref section**" which is not a standard LaTeX ref.
\hline
\end{tabular}
\end{table*}

This technical implementation of ANS ensures that universal agent Registry/Directory operates with cryptographic assurance, distributed consensus for critical operations, and real-time compliance enforcement while maintaining high performance and scalability requirements essential for enterprise AI agent deployments.

\section{Security Analysis and Considerations}
\label{sec:security}

\subsection{MAESTRO-Based Threat Analysis}
This section presents a systematic security threat analysis of the proposed ANS protocol. We identify key potential vulnerabilities and map them onto the MAESTRO 7 Layers framework \cite{mas_threat_model_2025} to provide a structured understanding of the threat landscape and the corresponding mitigation strategies integrated into our design. MAESTRO stands for Multi-Agent Environment, Security, Threat, Risk, and Outcome, and its seven layers are: Foundation Models, Data Operations, Agent Frameworks, Deployment and Infrastructure, Evaluation and Observability, Security and Compliance, and Agent Ecosystem. By analyzing vulnerabilities and risks at each architectural layer, as well as cross-layer interactions, MAESTRO enables security teams to proactively identify, assess, and mitigate threats unique to agentic AI.

\subsubsection{Threat: Agent Impersonation}
\begin{itemize}
    \item \textbf{Risk:} An adversary attempts to impersonate a legitimate, registered agent inside the ecosystem.
    \item \textbf{Mitigation Strategy:} Mandatory implementation of PKI. Verification of agent identity through validation of agent-specific digital certificates (\texttt{Cert}) issued by a trusted CA. Agent must prove possession of the private key.
    \item \textbf{Formal Check:}
        \begin{itemize}
            \item Agent MUST provide valid \texttt{Cert} on registration.
            \item RA MUST verify \texttt{Cert} against trusted CA.
            \item All communication MUST be digitally signed; recipient MUST verify signature using public key from \texttt{Cert}.
            \item See Section \ref{sec:pki_integration} (PKI Integration) and \ref{sec:agent_identity} (Agent Identity).
        \end{itemize}
    \item \textbf{MAESTRO Layer Mapping:} Layer 7 (Agent Ecosystem) - Agent Impersonation/Identity Attack.
\end{itemize}

\subsubsection{Threat: Registry Poisoning}
\begin{itemize}
    \item \textbf{Risk:} An adversary tries to inject malicious data into the Agent Registry (e.g., corrupting agentCapabilities or Endpoints).
    \item \textbf{Mitigation Strategy:} Strict RA validation during registration/renewal; cryptographic signing of registry responses (secure resolution); stringent database access controls \cite{Narajala2025ToolSquatting}. Secure resolution ensures responses are signed by the Agent Registry's private key, verifiable by clients.
    \item \textbf{Formal Check:}
        \begin{itemize}
            \item RA validation procedures (Section \ref{sec:registration_renewal}).
            \item Secure, authenticated registry responses (Section \ref{sec:ans_naming}.D leveraging PKI Section \ref{sec:pki_integration}).
            \item Database access controls (Infrastructure L4 / Data L2).
        \end{itemize}
    \item \textbf{MAESTRO Layer Mapping:}
        \begin{itemize}
            \item Layer 7 (Agent Ecosystem) - Compromised Agent Registry, Malicious Agent Discovery.
            \item Layer 6 (Security and Compliance) - RA validation policies, cryptographic signing.
            \item Layer 4 (Deployment and Infrastructure) / Layer 2 (Data Operations) - Database access controls.
        \end{itemize}
\end{itemize}

\subsubsection{Threat: Man-in-the-Middle (MitM) Attacks}
\begin{itemize}
    \item \textbf{Risk:} An adversary modifies communications between system components (agent-agent, agent-registry, agent-RA/CA).
    \item \textbf{Mitigation Strategy:} Message authenticity/integrity via digital signatures (agent's PKI private key). Secure transport (e.g., mTLS) is best practice but PKI signing is the core mechanism discussed. Formal resolution algorithm (\ref{sec:ans_naming}.C) enforces response integrity.
    \item \textbf{Formal Check:} PKI signing (Section \ref{sec:pki_integration}).
    \item \textbf{MAESTRO Layer Mapping:}
        \begin{itemize}
            \item Layer 4 (Deployment and Infrastructure) - Targets communication channels. Secure transport operates here.
            \item Layer 6 (Security and Compliance) - PKI framework providing keys/certs managed via L6.
        \end{itemize}
\end{itemize}

\subsubsection{Threat: Denial of Service (DoS) / Distributed Denial of Service (DDoS)}
\begin{itemize}
    \item \textbf{Risk:} Adversary attempts to incapacitate Agent Registry, RA, CA, or resolution services via traffic flooding or resource exhaustion.
    \item \textbf{Mitigation Strategy:} Resilience through distributed implementation design. Standard operational defenses (rate limiting, DDoS protection services). Formal resolution algorithms may include DoS limits.
    \item \textbf{Formal Check:} Architectural design (distributed options in Section \ref{sec:implementation}).
    \item \textbf{MAESTRO Layer Mapping:}
        \begin{itemize}
            \item Layer 4 (Deployment and Infrastructure) - DoS attacks target L4 availability. Architectural resilience is L4 design.
            \item Layer 7 (Agent Ecosystem) - Impact felt at L7 (disrupted discovery/interaction).
        \end{itemize}
\end{itemize}

\subsection{Additional Security Controls and Considerations}
Implementing ANS requires thorough threat modeling and comprehensive security controls.

\subsubsection{PKI Security Controls}
\begin{itemize}
    \item \textbf{Certificate Revocation:} Robust CRL \cite{RFC5280} or OCSP \cite{RFC6960} mechanisms are essential.
    \item \textbf{Secure Key Storage:} Agent private keys must be protected (HSMs, secure enclaves, OS key stores). Compromise requires immediate revocation/re-issuance.
    \item \textbf{Registry Access Control:} Strict authentication/authorization for registry management (RA) and potentially querying.
    \item \textbf{RA Validation:} Rigorous process to prevent malicious registration (domain validation, organizational checks, manual review).
\end{itemize}

\subsubsection{ANS-Specific Security Controls}
\begin{itemize}
    \item \textbf{Resolution Integrity:} Use DNSSEC-like mechanisms or signed responses to prevent ANS Manipulation/Spoofing.
    \item \textbf{DoS Mitigation:} Standard defenses (rate limiting, firewalls, anycast) for resolution Endpoints.
    \item \textbf{CA Security:} Standard CA best practices (offline roots, audits, HSMs) are mandatory. Consider using established, trusted CAs.
    \item \textbf{Sybil Attack Resistance:} Registration processes should make creating fake identities difficult/costly.
\end{itemize}

\subsubsection{Protocol Integration Security}
\begin{itemize}
    \item \textbf{Protocol-Specific Security:} Leverage security features within A2A, MCP, ACP (e.g., OAuth, capability tokens) for agent-to-agent communication post-discovery.
    \item \textbf{Governance and Trust Framework:} Clear policies, liability, and trust anchors are essential for adoption.
\end{itemize}

\subsubsection{Side-Channel Deanonymization and Mitigation}
\label{sec:side_channel_mitigation}
The process of querying the Agent Registry for specific agentCapabilities could unintentionally reveal sensitive information about the querier's intent, business objectives, or operational profile. To mitigate this risk, the implementation must prioritize:
\begin{itemize}
    \item \textbf{Private Information Retrieval (PIR) implementation:} To enable retrieval of information from the Agent Registry without revealing which information is being retrieved. Evaluate various PIR libraries to determine the most secure and efficient implementation based on use case. Prioritize the most performant and secure library.
    \item \textbf{Anonymized Query Relays:} Implementing query relays to hide the origin of the requests and prevent ANS Registry from seeing what agents are querying for what capabilities.
    \item \textbf{Differential Privacy for Aggregated Query Data:} Anonymize all data before sending to a 3rd party, ensuring that PII isn't included in the request.
    \item \textbf{Query Pattern Analysis:} Analyzing query patterns for anomalous behavior, and block queries that look suspicious.
    \item \textbf{Rate Limiting:} Implementing rate limiting to prevent potential DDoS attacks, and help obfuscate queries.
    \item \textbf{Auditing:} Ensure all security implementations are audited regularly.
    \item \textbf{Privacy Controls for Query and Response Data:} Implementation of data anonymization techniques to enhance privacy compliance.
\end{itemize}

\section{Implementation Considerations}
\label{sec:implementation}

The Agent Registry can be implemented using various patterns. The choice depends on scale, trust model, performance requirements, administrative overhead, and cost. The Protocol Adapter Layer suits a plugin architecture. Below is a decision matrix to help guide the selection:

\begin{table*}[!htbp]
\centering
\caption{Decision Matrix for Agent Registry Implementation Patterns}
\label{tab:implementation_decision_matrix}
\small
\begin{tabular}{|l|l|l|l|l|l|}
\hline
\textbf{Feature} & \textbf{Centralized} & \textbf{Distributed (Cassandra)} & \textbf{Distributed (DHT)} & \textbf{Blockchain/DLT} & \textbf{Federated} \\
\hline
Consistency & Strong & Tunable (Eventual) & Eventual & Strong & Tunable (Eventual) \\ \hline
Latency & Low & Medium & Medium to High & High & Medium to High \\ \hline
Scalability & Limited & High & Very High & Limited & High \\ \hline
Fault Tolerance & Low & High & High & High & Medium \\ \hline
Security & Medium & Medium & Medium & High & Medium \\ \hline
Operational Cost & Low & Medium & Medium & High & Medium \\ \hline
Complexity & Low & Medium & High & High & Medium \\
\hline
\end{tabular}
\end{table*}

\textbf{Implementation Patterns:}
\begin{itemize}
    \item \textbf{Centralized:} Simple management, but single point of failure/bottleneck. Suitable for smaller/private deployments.
    \item \textbf{Distributed (Cassandra, CockroachDB):} Offers resilience and scalability. Technologies like Cassandra provide tunable consistency (eventual). Requires coordination mechanisms.
    \item \textbf{Distributed (DHT):} Scalable P2P lookup, complex state management. Suitable for very large-scale deployments with relaxed consistency requirements.
    \item \textbf{Blockchain/DLT:} Smart contracts as Registry Agents, transactions validated and recorded on-chain. Offers high security and auditability but suffers from high latency, limited scalability, and significant write amplification. Write amplification refers to the fact that a single write operation to the blockchain can result in multiple write operations to the underlying storage, significantly increasing storage costs and reducing performance.
    \item \textbf{Federated:} Independent registries interoperate (organizational/geographic boundaries). Requires inter-registry protocols and trust. To enable seamless cross-registry discovery and interaction, federated registries require standardized mechanisms for verifying agent identities and resolving ANSNames across domains. This necessitates agreed-upon trust anchors (e.g., mutually trusted Certificate Authorities), standardized metadata formats for agent descriptions, and well-defined inter-registry communication protocols for querying and exchanging agent information. Without these elements, a federated system risks becoming a collection of isolated silos, hindering interoperability.
    \item \textbf{Hybrid:} Combines elements (e.g., centralized RA/CA, distributed/replicated read nodes). Caching layers (Redis, Memcached) improve performance. Requires careful consistency management.
\end{itemize}

\textbf{Key Considerations:}
\begin{itemize}
    \item \textbf{Consistency vs. Latency:} Strong consistency (e.g., in a centralized system) means that all reads see the most recent write, but it can increase latency. Eventual consistency (e.g., in Cassandra or a DHT) allows for lower latency and higher scalability, but reads may not always reflect the most recent write.
    \item \textbf{Write Amplification (Blockchain):} Be aware of the significant write amplification in blockchain solutions. This can lead to high storage costs and performance bottlenecks.
    \item \textbf{Operational Cost:} Consider the operational cost of each pattern, including hardware, software, and personnel costs. Blockchain solutions, in particular, can be expensive to operate due to the need for specialized hardware and expertise.
    \item \textbf{Complexity:} Distributed systems, especially those based on DHTs or blockchains, are more complex to design, implement, and maintain than centralized systems.
\end{itemize}

\section{Future Considerations}
\label{sec:future}

This proposal lays groundwork; future work should explore:
\begin{itemize}
    \item \textbf{Prototype Implementation:} Build and evaluate a working ANS prototype (See GitHub: \url{https://github.com/kenhuangus/dns-for-agents/}).
    \item \textbf{Explore an innovative bootstrap model for distributed trust infrastructure:} focusing on a foundation-led root governance approach with delegated sub-spaces (similar to Cloud Native Computing Foundation certificate transparency roots), alongside a comprehensive funding and fee schedule that creates a sustainable, flexible ecosystem for trust delegation and verification across different organizational scales and technological domains.
    \item \textbf{Performance \& Scalability:} Benchmark resolution latency, registration throughput, and scalability under load.
    \item \textbf{Advanced Cryptography:} Investigate privacy-preserving techniques (e.g., zero-knowledge proofs) for capability advertisement/selective disclosure.
    \item \textbf{Formal Verification:} Mathematically model and verify security properties of registration/resolution protocols.
    \item \textbf{Detailed Governance Model:} Develop comprehensive policies for naming, CA/RA operations, disputes, trust framework evolution.
    \item \textbf{Platform Integration:} Demonstrate integration with agent frameworks (LangChain, AutoGen, CrewAI) and cloud platforms.
    \item \textbf{Semantic Interoperability:} Research mechanisms leveraging ANS metadata for deeper cross-protocol communication (standardized capability ontologies, translation gateways).
    \item \textbf{Reputation Systems:} Integrate reputation scores/endorsements to enhance trust.
    \item \textbf{ANS agentCapability Negotiation and Binding Protocol:} Develop standardized protocols for agents to negotiate capabilities post-resolution (capability mapping taxonomies, binding, and quality of service).
    \item \textbf{Draft a governance white-paper} detailing name allocation, fee model, dispute arbitration, and root CA stewardship.
\end{itemize}

\section{Conclusion}
\label{sec:conclusion}

ANS offers a foundational infrastructure for a more secure, trustworthy, and interconnected agentic AI ecosystem. Its broad impact includes:
\begin{itemize}
    \item \textbf{Enhancing Interoperability:} Protocol-agnostic directory facilitates seamless communication between diverse agents.
    \item \textbf{Boosting Trust and Security:} PKI integration enhances trust for sensitive domains (finance, healthcare, cybersecurity).
    \item \textbf{Accelerating Innovation:} Lowers barriers by providing common discovery/identity, letting developers focus on agent solutions.
    \item \textbf{Facilitating Autonomous Systems:} Critical enabler for systems needing dynamic, secure discovery and interaction (autonomous vehicles, smart cities).
    \item \textbf{Powering Secure AI Marketplaces:} Foundation for marketplaces with verifiable agent identities/capabilities.
\end{itemize}
The modular Protocol Adapter Layer ensures extensibility. ANS provides foundational trust and discovery, enabling agents on different standards to find and securely contact each other. While challenges remain (governance, scalability, semantic interoperability), ANS is a necessary step towards a robust agentic AI ecosystem.

% Acknowledgements Section
\section*{Acknowledgment}
The authors acknowledge the contributions of the communities and organizations developing foundational agent communication standards, including Google (A2A), Anthropic (MCP), IBM (ACP), and the broader AI research community. That this is included in OWASP Gen AI Security Project, Agentic Security Initiative (ASI) - Agentic AI - Agent Name Service (ANS) for Secure AI Agent Discovery.

% Bibliography
\bibliographystyle{IEEEtran}
\bibliography{IEEEabrv,references} % Assumes references.bib contains your entries

% Generated by IEEEtran.bst, version: 1.14 (2015/08/26)
\begin{thebibliography}{10}
\providecommand{\url}[1]{#1}
\csname url@samestyle\endcsname
\providecommand{\newblock}{\relax}
\providecommand{\bibinfo}[2]{#2}
\providecommand{\BIBentrySTDinterwordspacing}{\spaceskip=0pt\relax}
\providecommand{\BIBentryALTinterwordstretchfactor}{4}
\providecommand{\BIBentryALTinterwordspacing}{\spaceskip=\fontdimen2\font plus
\BIBentryALTinterwordstretchfactor\fontdimen3\font minus \fontdimen4\font\relax}
\providecommand{\BIBforeignlanguage}[2]{{%
\expandafter\ifx\csname l@#1\endcsname\relax
\typeout{** WARNING: IEEEtran.bst: No hyphenation pattern has been}%
\typeout{** loaded for the language `#1'. Using the pattern for}%
\typeout{** the default language instead.}%
\else
\language=\csname l@#1\endcsname
\fi
#2}}
\providecommand{\BIBdecl}{\relax}
\BIBdecl

\bibitem{RFC1035}
\BIBentryALTinterwordspacing
P.~Mockapetris, ``Domain names - implementation and specification,'' RFC Editor, RFC 1035, Nov. 1987. [Online]. Available: \url{https://www.rfc-editor.org/info/rfc1035}
\BIBentrySTDinterwordspacing

\bibitem{RFC6763}
\BIBentryALTinterwordspacing
S.~Cheshire and M.~Krochmal, ``{DNS-Based Service Discovery},'' RFC Editor, RFC 6763, Feb. 2013. [Online]. Available: \url{https://www.rfc-editor.org/info/rfc6763}
\BIBentrySTDinterwordspacing

\bibitem{Surapaneni2025}
\BIBentryALTinterwordspacing
R.~Surapaneni, M.~Jha, M.~Vakoc, and T.~Segal, ``{Announcing the Agent2Agent Protocol (A2A)},'' Google for Developers Blog, Apr. 2025, accessed: 2025-04-27. [Online]. Available: \url{https://developers.googleblog.com/en/a2a-a-new-era-of-agent-interoperability/}
\BIBentrySTDinterwordspacing

\bibitem{A2A2025}
{Agent2Agent Protocol Specification Authors}, ``{Agent2Agent (A2A) Protocol Specification},'' \url{https://github.com/google/A2A}, 2025, accessed: 2025-04-27.

\bibitem{Anthropic2024}
Anthropic, ``Model context protocol ({MCP}),'' \url{https://www.anthropic.com/news/model-context-protocol}, 2024, accessed: 2025-04-27.

\bibitem{MCPSpec2025}
{Model Context Protocol Specification Authors}, ``{Model Context Protocol (MCP) Specification},'' \url{https://modelcontextprotocol.io/specification/2025-03-26}, Mar. 2025, accessed: 2025-04-27.

\bibitem{Schmid2025}
P.~Schmid, ``{MCP Introduction},'' \url{https://www.philschmid.de/mcp-introduction}, 2025, accessed: 2025-04-27.

\bibitem{IBM2025}
{IBM Research}, ``(placeholder for anticipated ibm acp publication/specification link),'' 2025, anticipated Publication.

\bibitem{FIPA2002}
\BIBentryALTinterwordspacing
{Foundation for Intelligent Physical Agents (FIPA)}, ``{FIPA Agent Communication Language Specifications},'' FIPA, Tech. Rep., 2002. [Online]. Available: \url{http://www.fipa.org/repository/aclspecs.html}
\BIBentrySTDinterwordspacing

\bibitem{RFC5280}
\BIBentryALTinterwordspacing
D.~Cooper, S.~Santesson, S.~Farrell, S.~Boeyen, R.~Housley, and W.~Polk, ``Internet x.509 public key infrastructure certificate and certificate revocation list (crl) profile,'' RFC Editor, RFC 5280, May 2008. [Online]. Available: \url{https://www.rfc-editor.org/info/rfc5280}
\BIBentrySTDinterwordspacing

\bibitem{RFC7159}
\BIBentryALTinterwordspacing
E.~T.~Bray, ``The javascript object notation ({JSON}) data interchange format,'' RFC Editor, RFC 7159, Mar. 2014. [Online]. Available: \url{https://www.rfc-editor.org/info/rfc7159}
\BIBentrySTDinterwordspacing

\bibitem{RFC6960}
\BIBentryALTinterwordspacing
S.~Santesson, M.~Myers, R.~Ankney, A.~Malpani, S.~Galperin, and C.~Adams, ``{X.509 Internet Public Key Infrastructure Online Certificate Status Protocol - OCSP},'' RFC Editor, RFC 6960, Jun. 2013. [Online]. Available: \url{https://www.rfc-editor.org/info/rfc6960}
\BIBentrySTDinterwordspacing

\bibitem{securing_a2a}
\BIBentryALTinterwordspacing
I.~Habler, K.~Huang, V.~S. Narajala, and P.~Kulkarni, ``Building a secure agentic {AI} application leveraging {A2A} protocol,'' \emph{arXiv preprint arXiv.2504.16902}, 2025. [Online]. Available: \url{https://www.arxiv.org/abs/2504.16902}
\BIBentrySTDinterwordspacing

\bibitem{narajala2025enterprise}
\BIBentryALTinterwordspacing
V.~S. Narajala and I.~Habler, ``{Enterprise-Grade Security for the Model Context Protocol (MCP): Frameworks and Mitigation Strategies},'' \emph{arXiv preprint arXiv:2504.08623}, 2025. [Online]. Available: \url{https://arxiv.org/abs/2504.08623}
\BIBentrySTDinterwordspacing

\bibitem{mas_threat_model_2025}
\BIBentryALTinterwordspacing
K.~Huang, A.~Sheriff, J.~Sotiropoulos, R.~F. Del, and V.~Lu, ``Multi-agentic system threat modelling guide {OWASP} {GenAI} security project,'' Apr. 2025. [Online]. Available: \url{https://www.researchgate.net/publication/391204915_Multi-Agentic_system_Threat_Modelling_Guide_OWASP_GenAI_Security_Project}
\BIBentrySTDinterwordspacing

\bibitem{Narajala2025ToolSquatting}
\BIBentryALTinterwordspacing
V.~S. Narajala, K.~Huang, and I.~Habler, ``Securing genai multi-agent systems against tool squatting: A zero trust registry-based approach,'' arXiv.org, 2025. [Online]. Available: \url{https://arxiv.org/abs/2504.19951}
\BIBentrySTDinterwordspacing

\end{thebibliography}

% Appendices
\appendices

\section{Complete Request/Response Schemas}
\label{app:schemas}
The detailed JSON Schema documents for registry interactions are maintained externally. Please ensure these schemas are well-commented and validated in any implementation.
\begin{itemize}
    \item AgentRegistrationRequest Schema: \url{https://github.com/kenhuangus/dns-for-agents/blob/main/agent_registration_request_schema.json}
    \item AgentRenewalRequest Schema: \url{https://github.com/kenhuangus/dns-for-agents/blob/main/agent_renewal_request_schema.json}
    \item AgentRegistrationResponse Schema: \url{https://github.com/kenhuangus/dns-for-agents/blob/main/agent_registration_response_schema.json}
    \item AgentRenewalResponse Schema: \url{https://github.com/kenhuangus/dns-for-agents/blob/main/agent_renewal_response_schema.json}
    \item AgentCapabilityRequest Schema: \url{https://github.com/kenhuangus/dns-for-agents/blob/main/agent_capability_request.schema.json}
    \item AgentCapabilityResponse Schema: \url{https://github.com/kenhuangus/dns-for-agents/blob/main/agent_capability_response.schema.json}
\end{itemize}

\section{Glossary of Terms - Agent Name Service (ANS)}
\label{app:glossary}

\textbf{A2A (Agent2Agent Protocol):}\par A communication protocol developed by Google for standardizing inter-agent communication, designed to bridge different agent frameworks.
\par\medskip

\textbf{ACP (Agent Communication Protocol):}\par A protocol designed by IBM Research to standardize how agents communicate, enabling automation, collaboration, UI integration, and developer tooling.
\par\medskip

\textbf{Agent:}\par An autonomous software entity capable of performing tasks, making decisions, and interacting with other agents or systems.
\par\medskip

\textbf{Agent Identity:}\par The verifiable identity of an agent within the ANS ecosystem, comprising cryptographic identity (PKI certificate), logical identity (ANSName), and protocol-specific identities.
\par\medskip

\textbf{Agent Registry:}\par A database storing registered agent information including capabilities, security policies, PKI certificates, protocol-specific metadata, and registration/renewal timestamps.
\par\medskip

\textbf{agentCapability:}\par A specific function, service, or skill that an agent can perform or provide to other agents or users.
\par\medskip

\textbf{ANS (Agent Name Service):}\par A universal directory service framework that enables secure discovery and interoperability between AI agents across different protocols and platforms.
\par\medskip

\textbf{ANSName:}\par A structured identifier for agents in the ANS ecosystem. 
\par\medskip

\textbf{CA (Certificate Authority):}\par A trusted entity that issues and manages digital certificates that bind public keys to entities (like agents) to establish a chain of trust in the ANS ecosystem.
\par\medskip

\textbf{Certificate Chain Verification:}\par The process of validating a certificate by checking the chain of trust from the certificate up to a trusted root certificate authority.
\par\medskip

\textbf{Certificate Revocation:}\par The process of invalidating a certificate before its expiration date, typically due to a key compromise or when an agent is deregistered.
\par\medskip

\textbf{Certificate Revocation List (CRL):}\par A list of digital certificates that have been revoked before their scheduled expiration date and should no longer be trusted.
\par\medskip

\textbf{CRL (Certificate Revocation List):}\par A mechanism used to check if a certificate has been revoked and is no longer valid.
\par\medskip

\textbf{CSR (Certificate Signing Request):}\par A message sent by an agent to a Certificate Authority to apply for a digital certificate.
\par\medskip

\textbf{Digital Signature:}\par A mathematical scheme for verifying the authenticity and integrity of digital messages or documents.
\par\medskip

\textbf{Distributed Hash Table (DHT):}\par A decentralized distributed system that provides a lookup service similar to a hash table, used as one possible implementation strategy for the Agent Registry.
\par\medskip

\textbf{DNS (Domain Name System):}\par The traditional system that translates human-readable domain names to IP addresses, which serves as a partial model for ANS but lacks the capability-oriented nature of ANS.
\par\medskip

\textbf{DNS-SD (DNS-Based Service Discovery):}\par An extension to DNS that enables automatic discovery of services available on a local network.
\par\medskip

\textbf{Endpoint:}\par A resolvable network address, service binding, or metadata document that allows agents to connect and communicate with each other.
\par\medskip

\textbf{Extension:}\par A field in the ANSName that holds deployment-specific or provider-defined metadata.
\par\medskip

\textbf{Interoperability:}\par The ability of different agent systems or protocols to exchange information and use that information effectively across platforms.
\par\medskip

\textbf{MAESTRO (7 Layers):}\par A threat modeling framework for agentic AI consisting of 7 layers, used to structure the security analysis of ANS.
\par\medskip

\textbf{MAS (Multi-Agent Systems):}\par Systems composed of multiple interacting intelligent agents that can cooperate, coordinate, or compete to solve problems.
\par\medskip

\textbf{MCP (Model Context Protocol):}\par A protocol developed by Anthropic focused on simplifying the integration of AI models with external tools and data sources.
\par\medskip

\textbf{OCSP (Online Certificate Status Protocol):}\par An internet protocol used for obtaining the revocation status of X.509 digital certificates as an alternative to CRLs.
\par\medskip

\textbf{PKI (Public Key Infrastructure):}\par A set of roles, policies, hardware, software, and procedures needed to create, manage, distribute, use, store, and revoke digital certificates and manage public-key encryption.
\par\medskip

\textbf{Protocol Adapter Layer:}\par A component in the ANS architecture that translates between the registry's internal representation and protocol-specific formats.
\par\medskip

\textbf{protocolExtensions:}\par A field within the ANS schema that acts as a container for protocol-specific data.
\par\medskip

\textbf{Provider:}\par An organization or entity that offers or maintains agents in the ANS ecosystem.
\par\medskip

\textbf{RA (Registration Authority):}\par An entity that verifies agent registration and renewal requests, interacts with the CA to issue certificates, and manages the agent lifecycle.
\par\medskip

\textbf{Registry Poisoning:}\par A security threat where an adversary attempts to inject malicious data into the Agent Registry.
\par\medskip

\textbf{Semantic Versioning:}\par A versioning scheme with a format of MAJOR.MINOR.PATCH used to indicate compatibility and changes in agent versions.
\par\medskip

\textbf{Service Discovery:}\par The process of automatically finding available services in a network.
\par\medskip

\textbf{Signature Verification:}\par The process of checking that a digital signature is valid and was created by the claimed signer.
\par\medskip

\textbf{Sybil Attack:}\par An attack where a malicious actor creates multiple fake identities to gain disproportionate influence in a network.
\par\medskip

\textbf{Trust Anchor:}\par The root of trust in a PKI system, typically a certificate authority whose certificate is implicitly trusted.
\par\medskip

\textbf{Version Negotiation:}\par The process where agents determine which protocol version to use when communicating, based on compatibility and preferences.
\par\medskip

\textbf{VerifyCertChain:}\par A function that checks the validity of a certificate by tracing its chain of trust back to a trusted certificate authority.
\par\medskip

\textbf{VerifySignature:}\par A function that validates a digital signature against a message and a public key to confirm authenticity and integrity.

\end{document}